%% file: manuscript.tex
\documentclass[aps,pra,twocolumn,superscriptaddress]{revtex4-2}
\usepackage[mathlines]{lineno}
\usepackage[scr=dutchcal]{mathalpha}
\usepackage{amsmath,bm}
\usepackage{amssymb}
\usepackage{physics}
\usepackage{graphicx}
\usepackage[percent]{overpic}
\begin{document}


\title{Topological Edge States and Collective Radiation in a One-Dimensional Atomic Chain}


\author{Arda Deniz İyican}
\email[]{ardaiyican@iyte.edu.tr}
\affiliation{Department of Physics, İzmir Institute of Technology, İzmir, Türkiye}
\author{Ahmet Levent Subaşı}
\email[]{alsubasi@itu.edu.tr}
\affiliation{Department of Physics, Istanbul Technical University, Istanbul, Türkiye}
\author{Özgür Çakır}
\email[]{ozgurcakir@iyte.edu.tr}
\affiliation{Department of Physics, İzmir Institute of Technology, İzmir, Türkiye}


\date{\today}

\begin{abstract}
We investigate the topological and collective radiative properties of a one-dimensional diatomic chain of identical quantum emitters with alternating spacing, coupled to the electromagnetic vacuum. The system realizes an extended, non-Hermitian Su–Schrieffer–Heeger–type model with vacuum-mediated long-range interactions and collective dissipation. We focus on the single-excitation manifold, where the dynamics are described by an effective Hamiltonian. For an infinite chain, the complex band structure reveals subradiant modes associated with wave vectors outside the light line, as well as parameter regimes where real-part band crossings occur. The bulk topology is characterized by a complex Berry phase, which remains quantized in the presence of inversion symmetry and a spectral gap. For finite chains, using exact diagonalization, we identify superradiant, subradiant, and edge states with distinct decay rates and spatial profiles. Edge states emerge in the topologically nontrivial regime when a bulk gap in the real part of the energy spectrum is present, while subradiant states exhibit strongly suppressed decay with system-size dependence. Finally, we analyze the far-field radiation patterns associated with different classes of eigenstates and show that the emission characteristics reflect the decay properties, spatial localization, and the parity of the unit-cell amplitudes.
\end{abstract}


\maketitle


\input{intro.tex}
\input{model.tex}
\input{periodicbc.tex}
\input{openbc.tex}
\input{radiation.tex}
\input{conclusion}

\appendix
\input{lerchtranscendent}
\input{nearfield}

\begin{acknowledgments}
A.L.S. would like to thank Kanupriya Sinha for useful discussions.
\end{acknowledgments}

\clearpage
\bibliography{bibliography}
\end{document}

%% file: intro.tex
\section{\label{introduction}Introduction}
Ensembles of closely spaced quantum emitters interacting with a common environment are known to form collective excitation states with modified relaxation dynamics, including collective decay rates that are enhanced (superradiance) or suppressed (subradiance) compared to those of individual emitters. This phenomenon was first pointed out by Dicke in his seminal work\cite{Dicke1954}, where he demonstrated enhanced spontaneous emission in a gas of emitters confined to a volume with dimensions smaller than the radiation wavelength.  He referred to the collective excitation states with enhanced emission as \textit{superradiant states} of the gas, in which the atomic cloud behaves as a collective dipole, resulting in radiation intensity that scales as $\sim N^2$ for $N$ atoms\cite{Dicke1954,Rehler1971, Lehmberg1970,Gross1982}. The innate counterparts of superradiant states are  \textit{subradiant states}, which arise from destructive interference between the emitters and lead to reduced collective decay rates \cite{Dicke1954, Gross1982}. While superradiance was observed by Gross et al. \cite{Gross1976} , the direct observation of subradiance has been more elusive because subradiant states couple weakly to the environment  \cite{Pavolini1985}. With the advent of atom trapping/cooling techniques, both superradiance and subradiance have been observed in trapped two-ion system \cite{Devoe1996}, cold atom ensembles \cite{Guerin2016, Ferioli2021, Ferioli2021dr}, and in superconducting qubits \cite{Wang2020}. Periodicity in ordered lattices of emitters allows collective effects to survive outside the regime where the system size can extend well beyond the transition wavelength. Such effects were predicted for atomic arrays in free space coupled via the electromagnetic vacuum\cite{Asenjo-Garcia2017,Zhang2020Scaling,Zhang2019,Zhang2022,Masson2020many}, and in the  context of waveguide QED, since waveguide mediated interactions between emitters survive beyond subwavelength limit \cite{Corzo2019, Sheremet2023,Tiranov2023,Solano2017,VanLoo2013,McDonnell2022subradiantedge,Zhang2020waveg}. The subradiant behavior of atomic ensembles may offer many practical applications such as efficient photon storage \cite{Ferioli2021,Facchinetti2016,Kalachev2007,Asenjo-Garcia2017,Cech2023,Ballantine2021,yelin2022}, high reflectivity atomic mirrors \cite{Rui2020}, atomic clocks with enhanced stability \cite{Henriet2019}. One-dimensional (1D) arrays of atoms can also be employed as atomic waveguides, where the subradiant states serve as guided modes by prohibiting decay into free space \cite{Masson2020waveguide,Asenjo-Garcia2019}. These collective states with significantly reduced coupling to the environment, leading to prolonged coherence times, effectively form decoherence-free subspaces, enabling deterministic generation of entangled states \cite{Ficek2002,bigorda2025,guimond2019,Gonzlez-Tudela2015}.

Topological phases of matter offer a platform for implementing efficient quantum technologies such as single photon generation with enhanced indistinguishability \cite{Wang2024}, fast and robust quantum-state transfer \cite{Lang2017,Dangelis2020}, 
Diatomic 1D arrays with mediated interactions differ from the standard SSH model \cite{Su1979} by inherent long range interactions within the chain, and non-Hermiticity due to dissipation.  Topological phenomena in systems with collective excitations have attracted significant interest \cite{Bettles2017,Perczel2017w,Perczel2017,Zhang2020Scaling,McDonnell2022subradiantedge}, since in such systems the mediated coherent interactions extend far beyond the commonly employed nearest-neighbor couplings, and their non-Hermitian nature invalidates the conventional bulk-boundary correspondence, requiring alternative formulations of topological invariants \cite{Lieu2018,Kunst2018,Liang2013,Ghatak2019,Bergholtz2021,Yokomizo2019,Zhang2025}.
The topological and collective properties of extended, non-Hermitian SSH-like models have been studied in several related systems. In plasmonic chains, the role of long-range dipole-dipole interactions, retardation, and radiative losses has been investigated in connection with topological edge modes and bulk-edge correspondence  \cite{Pocock2018,Pocock2019}. Similar questions have also been considered for quantum emitters coupled to a one-dimensional waveguide, where waveguide-mediated interactions lead to topological edge states with strongly suppressed decay rates \cite{McDonnell2022subradiantedge}. In dimerized atomic arrays implementing the SSH Hamiltonian, the topological transition, associated with the closing and reopening of the band gap, has been shown to change the scaling of subradiant decay rates \cite{Zhang2020Scaling}. 

In this work, we investigate the topological and subradiative behavior of a linear chain of identical two-level atoms with alternating spacings, coupled to a three-dimensional (3D) electromagnetic vacuum. The model is further described in Sec.~\ref{model}. Collective and topological properties are discussed in Sec.~\ref{infinite} and Sec.~\ref{finite} for infinitely extended, and finite atomic chains respectively. Finally, the behavior of radiation emitted from most subradiant, radiant and edge states as well as the properties of collective photon states formed by destructive or constructive dipole interference, are discussed in  Sec.~\ref{Radiation}.

%% file: model.tex
\section{\label{model}The Model}
We consider a one-dimensional linear array of identical two-level atoms in free space, with alternating interatomic spacings, forming a diatomic unit cell. Here the unit cell consists of two identical atoms as illustrated in Fig.~\ref{fig:modelandinfinitechainbands}(b). The interaction of atoms with the vacuum field is described by the electric-dipole interaction, within the rotating-wave approximation. Under the Born-Markov approximation, the system dynamics can be described by the Gorini-Kossakowski-Sudarshan-Lindblad (GKSL) master equation\cite{Barnett1997,Petruccionebook,Manzano2020}:
\begin{eqnarray}\label{gksl}
\dot{\rho}_{s} = -\frac{i}{\hbar}[ \mathcal{H},\rho_s(t)] + \mathscr{D}(\rho_s(t))
\end{eqnarray}
where, $\rho_s$ is the reduced density operator of the atomic system. The superoperator $\mathscr{D}\left(\rho_s(t)\right)$ which describes the collective decay of the system, is given by
\begin{eqnarray}\label{Dissipator} 
\mathscr{D}(\rho_s) = \sum_{\substack{n,m \\ \alpha,\beta}} \Gamma_{n\alpha;m\beta}\left[\sigma_-^{n\alpha}\rho_s \sigma_+^{m\beta} - \frac{1}{2} 
\{\sigma_+^{m\beta} \sigma_-^{n\alpha} , \rho_s \}\right].
\end{eqnarray} 
 Here, $\sigma_+^{n\alpha}$ and $\sigma_-^{n\alpha}$ are the raising and lowering operators for the atom in $n$-th unit cell and sublattice $\alpha=A,B$.  The term $\mathcal{H}$ in  Eq.\eqref{gksl} includes collective coherent interactions between the electric dipoles mediated by the vacuum:
\begin{eqnarray}\label{lambshift}
\mathcal{H} = \sum_{\substack{n\neq m\\ \alpha,\beta}} \Omega_{n\alpha;m\beta} \sigma_+^{m\beta} \sigma_-^{n\alpha}.
\end{eqnarray}
In Eqs. \eqref{Dissipator} and \eqref{lambshift}, $\Gamma_{n\alpha;m\beta}$ is the dissipative collective decay rate, and $\Omega_{n\alpha;m\beta}$ represents the coherent dipole-dipole interaction strength.
These collective parameters are given by the real and imaginary parts of the Green's function\cite{ficekbook,Ficek2002}:
\begin{eqnarray}
    {\cal G}(\mathbf r_{n\alpha;m\beta}) = \Omega_{n\alpha;m\beta} - \frac{i}{2} \Gamma_{n\alpha;m\beta}, \quad (n\neq m)
\end{eqnarray}
where 
\begin{widetext}
\begin{eqnarray} \label{greensR}
\begin{aligned} 
    {\cal G}({\bf R}) = \frac{3\Gamma_0}{4} \Biggl\{-\left[ 1- (\hat{\mu} \cdot \hat{R})^2 \right] \frac{1}{k_0 R}  + \left[1-3(\hat{\mu} \cdot \hat{R})^2\right]  \left( -\frac{i}{(k_0 R)^2}
     +\frac{1}{(k_0 R)^3}\right) \Biggl\} e^{ik_0 R}.
\end{aligned}
\end{eqnarray}
\end{widetext}
Here ${\bf r}_{n\alpha;m\beta} = {\bf r}_{n\alpha} - {\bf r}_{m\beta}$ is the interatomic distance. The positions of the atoms in the linear chain is given as $r_{nA} = na$ and $r_{nB} = na + b$, where $a$ is the lattice constant, and $b$ is the intracell separation.
$\Gamma_{n\alpha;n\alpha}=\Gamma_0 = \mu^2 \omega_0^3/(3 \varepsilon_0 \pi c^3 \hbar)$ is the single atom decay rate, where $\omega_0$ denotes the resonance frequency, and $\bm{\mu}$ is the transition dipole moment. All atoms forming the chain have the same dipole matrix elements ${\bf d}^{n\alpha} = {\bm{\mu}} {\hat{\sigma}_{-}}^{n\alpha} + \text{h.c.} $, and where dipole moments are all parallel, making an angle $\theta$ with the chain as illustrated in Fig.~\ref{fig:modelandinfinitechainbands}(b). 

As can be seen in Eq.~\eqref{greensR}, the coherent and dissipative dipole interactions depend polynomially on the interatomic distances. The coefficients of the terms with different power-law terms depend on the orientation of the dipole moments relative to the chain direction. In this work, we consider three distinct dipole orientations, corresponding to different power-law contributions of the Green’s function. In the first case, the transition dipole moments are aligned along the chain ($\theta = 0$), for which the coefficient of the $1/r$ term in Eq.~\eqref{greensR} vanishes, resulting in a faster decay of the interaction with distance. In the second case, the dipole moments are perpendicular to the chain ($\theta=\pi/2$). The final case corresponds to dipole moments forming an angle $\theta = \cos^{-1}(\sqrt{1/3})$ with the chain direction, for which the coefficients of the $1/r^{2}$ and $1/r^{3}$ contributions in Eq.~\eqref{greensR} vanish. Consequently, in this special configuration, the interaction decays as $1/r$. In general, the interaction decreases relatively slow with distance and extends well beyond nearest neighbor distances.

Further, the jump-free, non-unitary dissipation terms and the jump term in Eq.~\eqref{Dissipator} can be separated, allowing the master equation to be reformulated by introducing a non-Hermitian effective Hamiltonian that conserves excitation number and describes the continuous dynamics (without jump) of the system \cite{Carmichael1993,Molmer:93,plenio1998,Orszag2016,Daley2014} as follows;

\begin{eqnarray}
\dot{\rho}_s = -\frac{i}{\hbar}(H_{\rm eff}\rho_s-\rho_sH^\dagger_{\rm eff}) +  \sum_{\substack{n,m\\ \alpha,\beta}} \Gamma_{n\alpha;m\beta} \, \sigma_-^{n\alpha} \rho_s \sigma_+^{m\beta}
\end{eqnarray}
where the effective Hamiltonian is given by
\begin{eqnarray} \label{effectivehamiltonian}
\begin{aligned}
    H_{\rm eff} &=\mathcal{H} - \frac{i}{2} \sum_{\substack{n,m \\ \alpha,\beta}}  \Gamma_{n\alpha;m\beta}\,  \sigma_+^{n\alpha} \sigma_-^{m\beta}
    \\ 
    &=\sum_{\substack{n,m \\ \alpha,\beta}} {\cal G}({\bf r}_{n\alpha;m\beta}) \sigma_+^{n\alpha} \sigma_-^{m\beta}  
\end{aligned}
\end{eqnarray}
where the Hermitian Hamiltonian $\mathcal{H}$ introduced in Eq.\eqref{lambshift}, generates coherent unitary dynamics, while non-unitary evolution arises from the non-Hermiticity of $H_{\rm eff}$ due to the second part with dissipative rate $\Gamma_{n\alpha;m\beta}$. 

%% file: periodicbc.tex
\section{\label{infinite}Infinite Chain}
We first consider an infinite di-atomic linear chain with single collective excitation. Due to the translational invariance of the system, we employ Bloch’s theorem, allowing us to express the eigenstates of the effective system Hamiltonian as
\begin{eqnarray} \label{infinitechaininitialstate}
   \ket{\Psi(k)} = \frac{1}{\sqrt{N}}\sum_{n,\alpha} e^{ikna} u_{\alpha}(k) \ket{n,\alpha} 
\end{eqnarray}
where $\ket{n_{\alpha}} = \sigma_+^{n\alpha} \ket{g}^{\bigotimes 2N}$ denotes the state in which the atom at unit cell $n$ and sublattice $\alpha$ is excited, while all others remain in the ground state. The coefficients $u_{k\alpha}$ give amplitudes on the two sublattices and determine the corresponding Bloch bands. Within the single excitation manifold, the effective Hamiltonian in the Brillouin zone  is defined from the eigenvalue relation for the sub-lattice components of the Bloch states
\begin{eqnarray}\label{effectivehamiltoniank}
\underbrace{\begin{pmatrix}
G_{A A}(k) & G_{A B}(k)  \\
G_{B A}(k) & G_{B B}(k) \end{pmatrix}}_{H_{\rm eff}(k)} \begin{pmatrix}
    u_A(k)\\ u_B(k)
\end{pmatrix}=E \begin{pmatrix}
    u_A(k)\\ u_B(k)
\end{pmatrix}.
\end{eqnarray} 
The matrix elements of $H_{\rm eff}(k)$ are;
\begin{equation}\label{eq:Galphabeta}
G_{\alpha\beta}(k) =  \Omega_{\alpha\beta}(k)-\frac{i}{2}\Gamma_{\alpha\beta}(k)
\end{equation}
Where, $\Omega_{\alpha\beta}(k)$ and $\Gamma_{\alpha\beta}(k)$ are simply the Fourier transforms of the real space coherent and dissipative interactions. The open forms of $G_{\alpha \beta}(k)$ are given by
\begin{widetext}
\begin{subequations}\label{eq:openforms}
 \begin{eqnarray}
     G_{A A}(k) = G_{B B}(k) = \frac{3\Gamma_0}{4} \sum_{\substack{\ell \neq 0}} {\cal G}(\ell a) \, e^{ik \, \ell a}- \frac{i \Gamma_0}{2},
     \label{gaa}
\end{eqnarray}
\begin{equation}
     G_{A B}(k) = \frac{3\Gamma_0}{4} \sum_{\substack{\ell}} {\cal G}(|\ell \, a + b|) \, e^{ik\ell \, a },
     \label{gab}
\end{equation}
\begin{equation}
     G_{B A}(k) = \frac{3\Gamma_0}{4} \sum_{\substack{\ell}}  {\cal G}(|\ell \, a - b|)\, e^{ik\ell \, a}.
     \label{gba}
\end{equation}
\end{subequations}
\end{widetext}
Where, $\ell = n-m$. The infinite sums in Eq.~\eqref{eq:openforms} are in the form of Lerch Transcendent and the resulting expressions are represented using special functions as detailed in Appendix \ref{lerchtranscendent}. As also discussed in Appendix \ref{lerchtranscendent}, that for $\theta \neq 0$, the summation of $1/r$ term over $\ell$ diverges at $k=k_0$. Therefore, we introduce a  regularization factor $e^{-\varepsilon \abs{\ell}}$ (with $\varepsilon \to 0^{+}$) to guarantee the convergence of the series. 

\begin{figure*}[htbp]
    \centering
        \begin{overpic}[width=\linewidth]{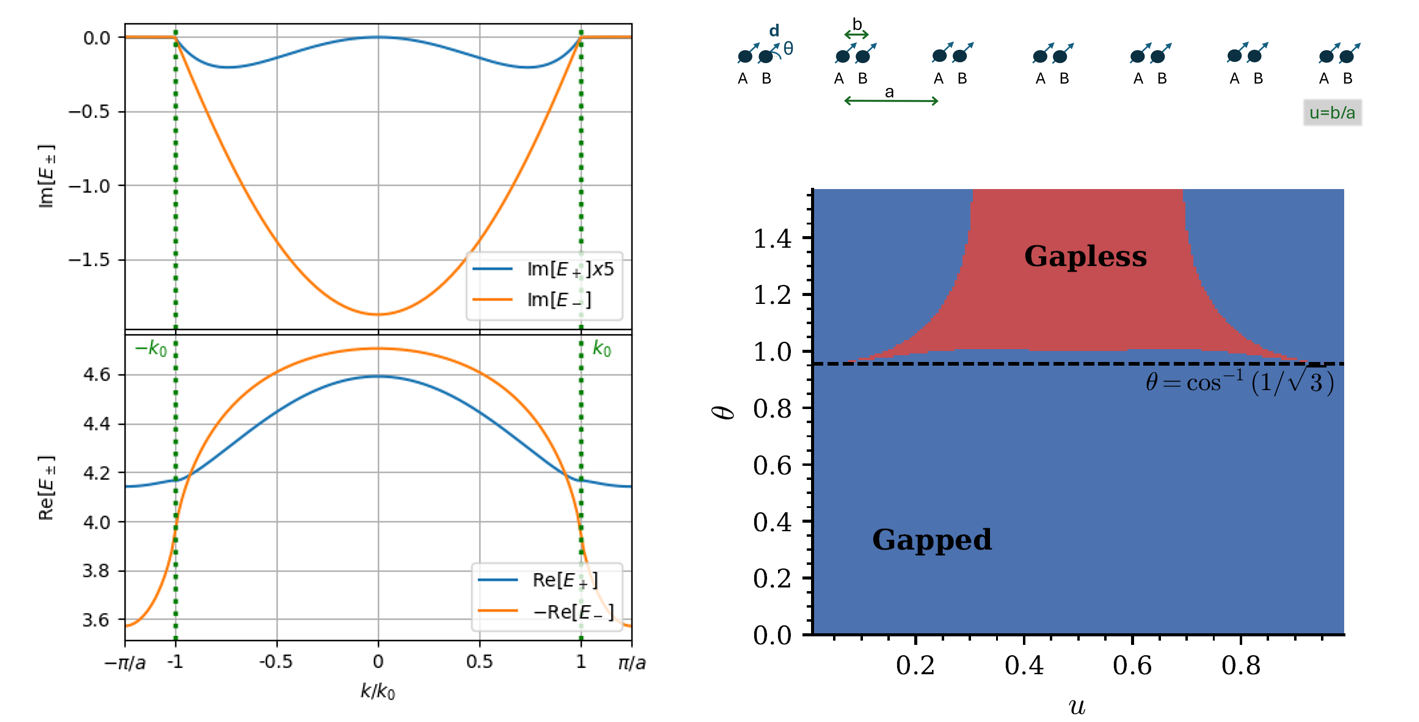}
            \put(1,49){\small\bfseries (a)}\put(48,49){\small\bfseries (b)}\put(48,37){\small\bfseries (c)}
        \end{overpic}
    \caption{(a) Real (bottom row) and imaginary (top row) components of the complex band structure for $a = 0.4 \, \lambda_0$, $u = 0.7$ and $\theta=0$. The real parts form the energy bands, while the imaginary parts determine the collective decay rates via $\Gamma_\pm(k) = -2 \, \Im[E_\pm(k)]$. The green dashed lines indicate the light lines $k=\pm k_0$. (b) Schematic of the model: a linear diatomic chain with lattice constant $a$, intracell separation $b$, ratio $u=b/a$, and dipole moments aligned at an angle $\theta$ with respect to the chain direction. (c) Phase diagram showing gapped/ungapped regions for $a=0.4\lambda_0$: The red region indicates parameter values $(\theta,u)$ where the real parts of the bands cross at some point in the Brillouin zone, while the blue region corresponds to a finite band gap. \iffalse (For convergence a value of $\epsilon = 10^{-7}$ was chosen.)\fi}
    \label{fig:modelandinfinitechainbands}
\end{figure*}

Given that $G_{\alpha\beta}(k)\neq G^*_{\beta\alpha}(k)$ and $G^*_{\alpha\alpha}\neq G_{\alpha\alpha}$, the effective Hamiltonian in Eq.~\eqref{effectivehamiltoniank} is non-Hermitian, and the resulting Bloch spectrum is generally complex. Moreover, since the atoms within the unit cell are identical, the diagonal elements that represent the interactions between the atoms in the same sublattices are equivalent for the two sublattices, such that $G_{AA}(k) = G_{BB}(k)$. Consequently, the two complex energy bands are 
\begin{align}\label{infiniteEngs}
    E_\pm(k) = G_{\alpha\alpha}(k) \pm \sqrt{G_{AB}(k) G_{BA}(k)}.
\end{align} 
The real parts, ${\rm Re}[E_\pm (k)]$, defines the band dispersion, whereas the imaginary part, $\Im[E_\pm (k)]$, encodes the decay rate of the excitation. In Fig.~\ref{fig:modelandinfinitechainbands}(a) the real and imaginary bands for singly excited infinite chain are shown for $a = 0.4 \, \lambda_0$. It can be seen from the imaginary part (upper panel) that the system enters a regime exhibiting subradiant behavior, where the excitation remains confined within the chain and does not decay into free space.  These dark states correspond to the off-resonant modes that lie outside the light line ($\abs{k}>\abs{k_0}$) and such modes fall within the Brillouin zone only when $a < 0.5 \, \lambda_0$. We can also see in the real bands in Fig.~\ref{fig:modelandinfinitechainbands}(a), that the modes outside the light line appear flatter, reflecting the weak dispersion of non-radiative modes that do not couple to free space. The observation of completely dark states in an infinite chain hinges on the lattice parameter being less than the critical value $\lambda_0/2$, rather than the atomic spacings in general. For $ a > 0.5 \, \lambda_0$, there are no dark states even though the nearest neighbour atomic spacing is less than half the transition wavelength since what allows the destructive interference to survive beyond Dicke regime is the spatial periodicity. 

While the global band gap closes at $u=0.5$, we also observe that for particular values of $u$ and $\theta$, the real parts of the two bands intersect at isolated points within the Brillouin zone.
The diagram in Fig.~\ref{fig:modelandinfinitechainbands}(c) shows the regions in the $(u,\theta)$ parameter space where real-band crossings occur. It indicates that the bands remain gapped for all values of $u$ when $\theta \textless \arccos(1/\sqrt{3})$. In particular, the condition $\Re[E_+] = \Re[E_-]$ requires $\Re\left[(G_{AB}(k) G_{BA}(k))^{1/2}\right] = 0$ in Eq. \eqref{infiniteEngs}. This is satisfied when ${\rm Im}\left[G_{AB}(k) G_{BA}(k)\right] = 0$ and  ${\rm Re}\left[G_{AB}(k) G_{BA}(k)\right] \leq 0$. This, in turn, corresponds to the regime $1-3 \cos^2\theta \, \textgreater \, 0$, i.e., $\theta \textgreater \arccos(1/\sqrt{3})$ as can be inferred from  Eqs.~\eqref{eq:openforms} and \eqref{greensR} where the imaginary part of $G_{AB}(k)G_{BA}(k)$ vanishes. Since this band crossing occurs only in the real part of the spectrum and does not correspond to a full complex degeneracy, the geometric phase remains well defined under periodic boundary conditions despite the real-part band touching. However, in the absence of a spectral gap, the bulk invariant no longer guarantees the existence of robust edge states under open boundary conditions. This issue will be discussed in Sec.~\ref{finite}.

\subsection{Topological Properties of The Atomic Chain}

As a starting point for analyzing the bulk topology of the atomic chain, we rewrite the two-band effective Hamiltonian in Eq.~\eqref{effectivehamiltoniank} in terms of Pauli matrices.
\begin{eqnarray} \label{hamiltoniantopology}
H_{\rm eff}(k) =
\begin{pmatrix}
G_{\alpha \alpha} & \gamma_x - i\gamma_y & \\
\gamma_x + i\gamma_y & G_{\alpha\alpha}& \end{pmatrix} = G_{\alpha\alpha} \, \sigma_{0} + \bm{\gamma} \cdot \bm{\sigma}.
\end{eqnarray}
As mentioned above, the diagonal terms are equal to each other; $G_{\alpha\alpha} = G_{AA} = G_{BB}$ and are shown in Eq.~\eqref{gaa}. The components $\sigma_i$ of the vector $\bm{\sigma}$ are usual Pauli matrices where $i=x,y,z$ and the components $\gamma_i$ of the Bloch vector are complex functions of $k$, with $\gamma_x = (1/2)(G_{AB} + G_{BA})$ and $\gamma_y = (i/2)(G_{AB}-G_{BA})$. It can be revealed from the open forms of $G_{\alpha\beta}(k)$ shown in Eqs. \eqref{gab} and \eqref{gba}, that the real and imaginary parts of the dissipative and coherent interactions are related to each other as;
\begin{eqnarray}
\begin{aligned}
    \Omega_{\alpha\beta}(k) = \Omega_{\beta\alpha}^*(k), \qquad  \Gamma_{\alpha\beta}(k) = \Gamma_{\beta\alpha}^*(k).
\end{aligned}
\end{eqnarray}
This gives, the components of the Bloch vector as $\gamma_x = \Re[\Omega_{AB}]-\frac{i}{2}\Re[\Gamma_{AB}]$, $\gamma_y = -\text{Im}[\Omega_{AB}]+\frac{i}{2}\text{Im}[\Gamma_{AB}]$, $\gamma_z = 0$. The two distinctions of our model from the standard SSH model \cite{Su1979} are that the effective Hamiltonian is non-Hermitian, and that long-range hoppings are allowed. It is well known that in one-dimensional systems, chiral symmetry ensures a quantized Zak phase \cite{zak1989}. This quantization plays a central role in characterizing topological phases, linking the geometric phase accumulated across the Brillouin zone to the emergence of edge states via the bulk–boundary correspondence. The standard Su–Schrieffer–Heeger (SSH) model belongs to the chiral orthogonal (BDI) class in the Altland–Zirnbauer classification when the hopping amplitudes are real, as it then possesses time-reversal, particle–hole, and chiral symmetries; for generic complex hopping amplitudes that break time-reversal symmetry, it reduces to the chiral unitary (AIII) class \cite{Altland1997,Ryu2010}. It allows only nearest-neighbor inter-sublattice hopping. Consequently, the Hamiltonian lacks a term proportional to $\sigma_z$ or $\sigma_0$, which results in $\sigma_z H \sigma_z^{-1} = -H$. In contrast, our system includes all-to-all hoppings, which trivially break chiral symmetry through a term in the Hamiltonian in Eq.~\eqref{hamiltoniantopology} that is proportional to the identity. However, since the second term in Eq.~\eqref{hamiltoniantopology} respects chiral symmetry, as $\gamma_z = 0$, 
and the identity term does not alter the eigenstates, the inversion symmetry of the model still guarantees a quantized Zak phase in our system. This is because the sublattices remain identical, and intra-sublattice hoppings do not contribute to a change in geometric phase. Therefore, when the topological invariant is nontrivial, the edge states exist and remain topologically protected by the bulk~\cite{longhi:18, Jiao2021,Pocock2019}.
As a consequence of non-Hermiticity, the Hamiltonian possesses complex spectrum and bi-orthogonal eigenvectors, which satisfy the relation $\bra{\psi^L_i}\ket{\psi_j^R} = \delta_{ij}$, where $\ket{\psi_i^R}$ are the right eigenstates of the Hamiltonian, and $\ket{\psi^L_i}$ are the eigenstates of its Hermitian conjugate. The right eigenstates of the effective Hamiltonian in Eq \eqref{hamiltoniantopology} are:

\begin{subequations}
\begin{equation}
\ket{\Psi_{\pm}^R} = 
\frac{1}{ \sqrt{2(\gamma_x^2+\gamma_y^2)}} 
\begin{pmatrix}
\gamma_x-i\gamma_y \\ 
\pm \sqrt{\gamma_x^2 + \gamma_y^2}
\end{pmatrix}
\end{equation}
With, eigenvalues, $E_{\pm} = G_{\alpha\alpha}(k)\pm \sqrt{\gamma_{x}^2 + \gamma_{y}^2}$. The left eigenvalues and eigenstates are the ones of its Hermitian conjugate $H_{\rm eff}^{\dagger}$:
\begin{eqnarray}
    \ket{\Psi_{\pm}^{L}} =\frac{1}{\sqrt{2(\gamma_{x}^2+\gamma_{y}^2)}^*}
\begin{pmatrix}
\gamma_{x}^*-i\gamma_{y}^* \\
\pm \sqrt{\gamma_{x}^2 + \gamma_{y}^2}^* 
\end{pmatrix}
\end{eqnarray}
\end{subequations}
The corresponding complex eigenvalues are $ E^{L}_{\pm} = G_{k}^{*}\pm \sqrt{\gamma_{x}^2 + \gamma_{y}^2}^{*}$. Since the normalisation is also defined in a biorthogonal sense, the geometric phase is defined for non Hermitian systems in biorthonormal basis as \cite{Lieu2018};
\begin{eqnarray}
 Q_{\pm} = i\int_{-\frac{\pi}{a}}^{\frac{\pi}{a}}
    \mel**{\Psi_{\pm}^{L}}{\partial_{k}}{\Psi_{\pm}^{R}} dk
\end{eqnarray}
This biorthogonal formulation of the geometric phase is referred to as the complex Berry phase. To proceed, $\ket{\Psi_{\pm}^{R}}$ and, $\ket{\Psi_{\pm}^{L}}$ can be parametrized as;
\begin{subequations}
    \begin{eqnarray}
    \ket{\Psi_{\pm}^{R}} =\frac{1}{\sqrt{2\, \text{cos}(\beta/2)\,\text{sin}(\beta/2)}}
\begin{pmatrix}
    \text{cos}(\beta/2)\\
\pm \, \text{sin}(\beta/2) \, e^{i\varphi} 
\end{pmatrix}
\end{eqnarray}

\begin{eqnarray}
    \ket{\Psi_{\pm}^{L}} =\frac{1}{\sqrt{2\, \text{cos}(\beta/2)\, \text{sin}(\beta/2)}}
\begin{pmatrix}
\text{sin}(\beta/2)\\
\pm \, \text{cos}(\beta/2) \, e^{i\varphi} 
\end{pmatrix}
\end{eqnarray}
\end{subequations}
where $\beta$ and $\varphi$ are defined as $\tan\beta = \abs{(\gamma_x +i\gamma_y)/\sqrt{\gamma_x^2 + \gamma_y^2}}$ and, the phase $\varphi = {\rm Arg}[\gamma_x +i\gamma_y]- {\rm Arg}[\sqrt{\gamma_x^2 + \gamma_y^2}]$. This yields, a real valued Berry phase;
\begin{eqnarray}
 Q_{\pm} = \frac{1}{2}\int_{-\frac{\pi}{a}}^{\frac{\pi}{a}}
   \Dot{\varphi} \, dk.
\end{eqnarray}
\begin{figure}[t]
    \centering
        \begin{overpic}[width=\linewidth]{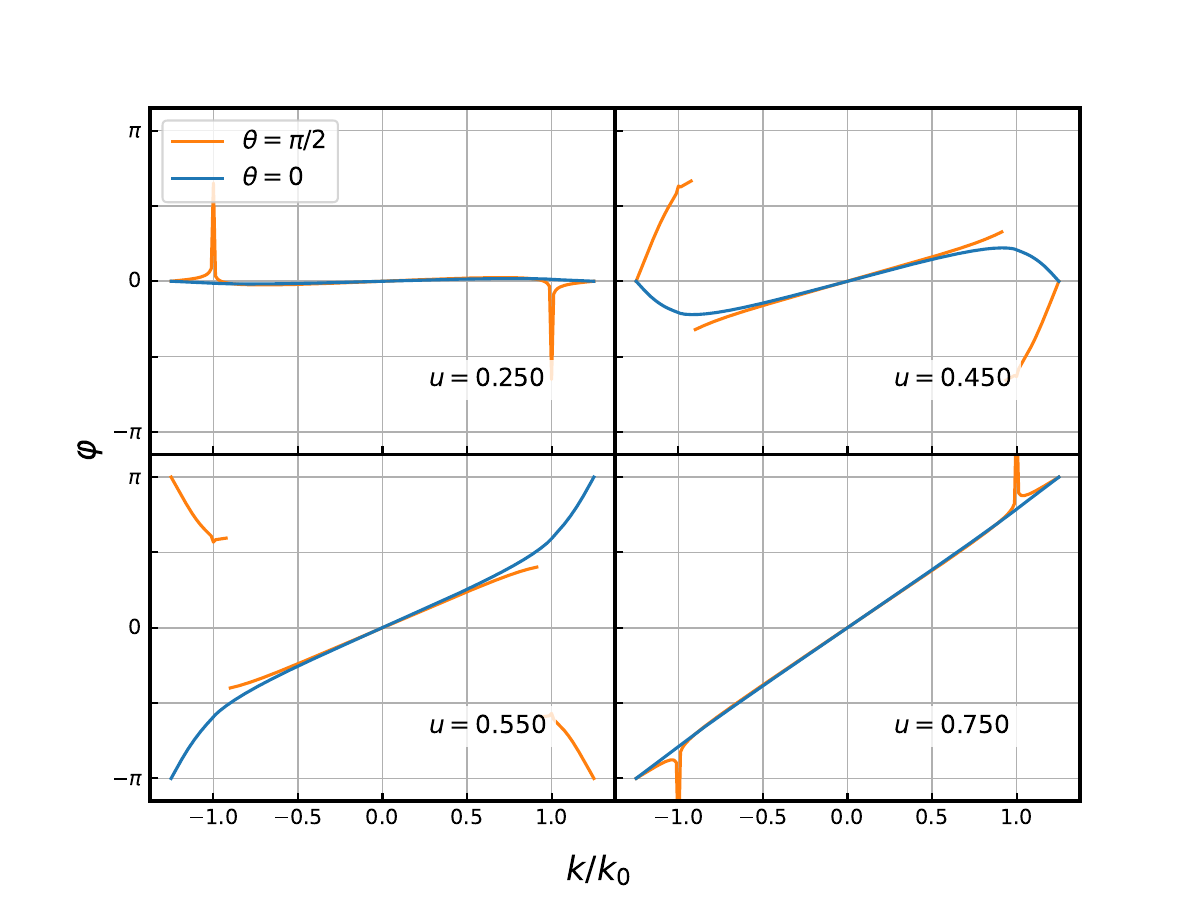}
        \put(43,62){\small\bfseries (a)}\put(80,62){\small\bfseries (b)}
    \put(80,33){\small\bfseries (d)} \put(43,33){\small\bfseries (c)}
        \end{overpic}
    \caption{Phase $\varphi$ across the Brillouin zone for an infinite chain with lattice spacing $a= 0.4 \lambda_0$, shown for different values of $u$. 
    }
    \label{fig:cberry}
\end{figure}
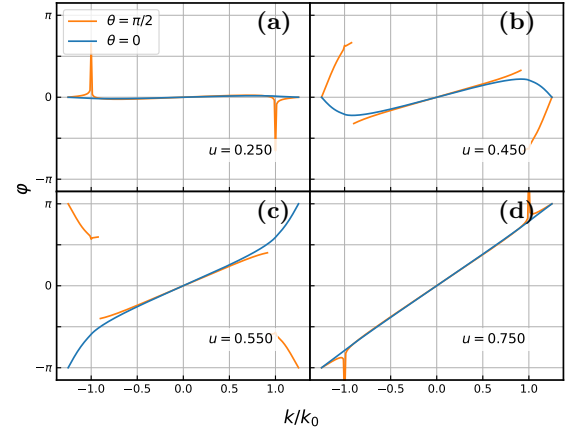
Which takes values $0, \, \pi \, \, (mod\,\pi)$ indicating trivial and non-trivial topology respectively. The phase $\varphi$ is plotted in the first Brillouin zone for different $u$ values when the orientation of dipole moments are parallel and perpendicular to the chain, in Fig.~\ref{fig:cberry}. As seen in the figure, for $\theta = 0$, when the intracell hopping is weaker than the intercell hopping ($u < 0.5$), the phase accumulated over the Brillouin zone is zero, while for $u > 0.5$, the state acquires a phase of $Q_\pm = \pi$ over a complete cycle, consistent with the Hermitian SSH model with nearest-neighbor hoppings \cite{Su1979}. Occurrence of non-trivial topology for $u > 0.5$ holds regardless of the orientation of the dipole moment as long as there is a finite band gap. However, at a real-band crossing, the phase of the eigenvector exhibits a discontinuous jump of $\pi$. Depending on how the phase evolves across the Brillouin zone, these discontinuities may cancel or accumulate to a total winding, corresponding to trivial or topological phases, respectively These behaviors are seen for $\theta = \pi/2$, $u = 0.450$ with $Q_\pm = \pi$ and $u = 0.550$ with $Q_\pm = 0$ in Fig.~\ref{fig:cberry}(b,c).

%% file: openbc.tex
\section{\label{finite}Finite Chain}
For a finite chain of $N$ unit cells, the real-space Hamiltonian in Eq. \eqref{effectivehamiltonian} has $2N$ biorthogonal eigenstates, indexed as $\ket{\Psi_\xi}$, with corresponding complex eigenvalues $E_\xi$, where $\xi = 1...2N$, and sorted according to the real parts of corresponding eigenvalues. These eigenstates can be represented as a superposition of local single-excitation states, each weighted by the amplitudes $v_{\xi;n\alpha}$ at the corresponding site,
\begin{equation}\label{finiteeigstate}
    \ket{\Psi_\xi} = \sum_{n\alpha} v_{\xi; n\alpha} \, \sigma^+_{n\alpha} \ket{g},
\end{equation}
\begin{figure*}[htbp]
    \centering
    \begin{minipage}[t]{0.32\textwidth}
        \begin{overpic}[width=\linewidth]{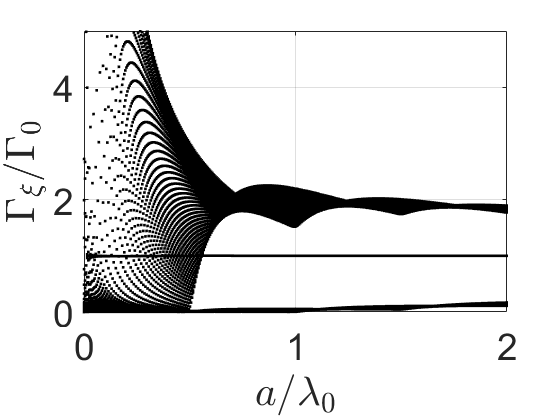}
            \put(0.5,65){\small\bfseries (a)}
        \end{overpic}
    \end{minipage}
    \hfill
    \begin{minipage}[t]{0.32\textwidth}
        \begin{overpic}[width=\linewidth]{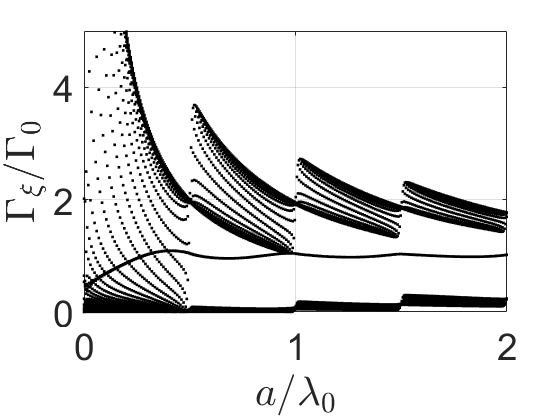}
            \put(0.7,65){\small\bfseries (b)}
        \end{overpic}
    \end{minipage}
    \hfill
    \begin{minipage}[t]{0.32\textwidth}
        \begin{overpic}[width=\linewidth]{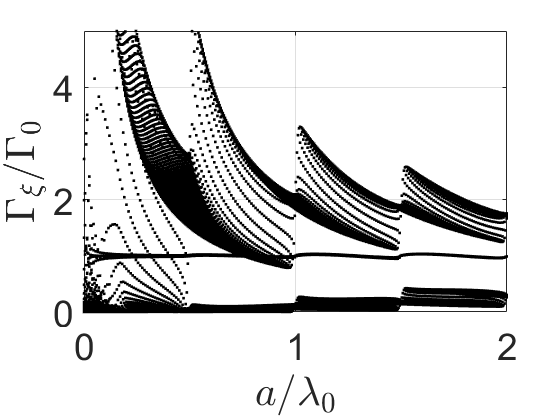}
            \put(0.7,65){\small\bfseries (c)}
        \end{overpic}
    \end{minipage}
    \caption{Plots of decay rates vs unit cell length for different dipole orientations (a) $\cos{\theta} = 1$, (b) $\cos{\theta} = 1/\sqrt{3}$, (c) $\cos{\theta} = 0 $. Here, the chain consists of $N = 50$ unit cells with intracell spacings $u = 0.9$. (The figures are zoomed in to make the details visible; several extremely superradiant states therefore lie outside the displayed range.) }
    \label{fig:decayratevsa}
\end{figure*}
with complex eigenvalues $E_\xi$, where $\text{Im}[E_\xi] = -\Gamma_\xi/2$ gives the decay rate, while the real part $\Re[E_\xi]$ corresponds to the energy shift from the atomic transition energy $\hbar \omega_0$. The eigenvalues and eigenstates are obtained by numerically diagonalizing $H_\text{eff}$. In Fig.~\ref{fig:decayratevsa}, the dependence of the decay rates on the cell size $a/\lambda_0$ is shown for the three dipole orientations with $u = 0.9$. In Fig.~\ref{fig:decayratevsa}(a-c), we see two bands, one with a relatively lower dispersion, and edge states due to non-trivial topology of the system. For all three angles, we see that collective subradiance is present for $a<0.5\lambda_0$ and for larger lattice constants, the collective decay rates converge towards single atom decay rate $\Gamma_0$. In contrast to the dark modes shown for an infinite linear chain in Fig.~\ref{fig:modelandinfinitechainbands}(a), the subradiant states for finite chain exhibit nonzero decay rates, although these are substantially smaller than $\Gamma_0$ depending on the system size. In Fig.~\ref{fig:NvsGamma}(a), the decay rates of the three most subradiant states versus the number of atoms forming the chain are plotted. The $1/N^3$ scaling observed in one-dimensional monatomic models persists in the diatomic chain \cite{Asenjo-Garcia2017,Zhang2020Scaling}.

The value of the intracell separation in Fig.~\ref{fig:decayratevsa} lies within the range where we expect the gap to be open for all angles $\theta$ [see Fig. \ref{fig:modelandinfinitechainbands}(c)], and the bulk topology is nontrivial. Thus, edge states are expected and observed as having decay rates close to the decay rate of a single isolated atom $\Gamma_{edge}/\Gamma_0 \approx 1$. While they are closest to $\Gamma_0$ for the interactions that decay faster with distance (Fig.~\ref{fig:decayratevsa}(a): $\theta = 0$), relatively stronger deviation is observed when only slowly decaying interactions are considered (Fig.~\ref{fig:decayratevsa}(b): $\theta=\cos^{-1}(1/\sqrt{3})$).
\begin{figure}[htbp]
    \centering
    \begin{minipage}[t]{0.25\textwidth}
        \begin{overpic}[width=\linewidth]{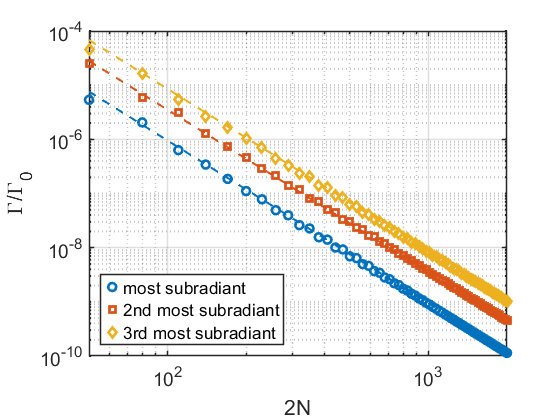}
            \put(0.5,70){\small\bfseries (a)}
        \end{overpic}
    \end{minipage}
    \hfill
    \begin{minipage}[t]{0.20\textwidth}
        \begin{overpic}[width=\linewidth]{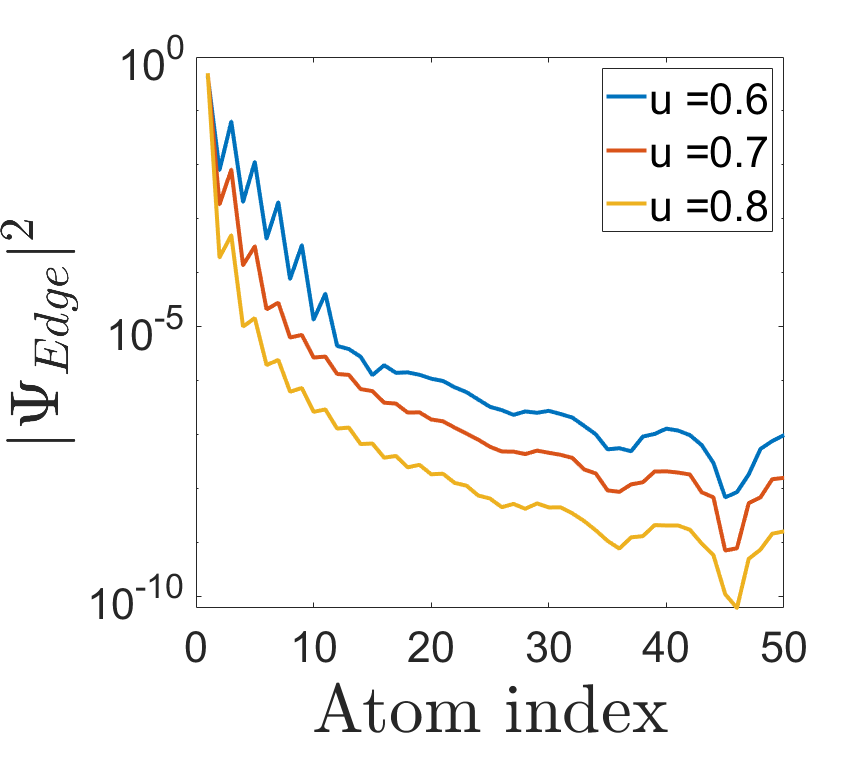}
            \put(0.7,87.5){\small\bfseries (b)}
        \end{overpic}
    \end{minipage}
    \caption{(a) Atom number dependence of the three minima of decay rates in logarithmic scale for $u=0.9$. Dashed lines are $(2N)^{-3}$ displayed for reference. (b) The probability distributions of one of the edge states of chain of $2N = 100$ atoms for various $u$ values in logarithmic scale. The other half of the chain gives mirror image of the shown. In both (a) and (b), $\theta=0$, $a=0.4 \, \lambda_0$}
    \label{fig:NvsGamma}
\end{figure}

\begin{figure*}

    \centering
    \begin{overpic}[width=\linewidth]{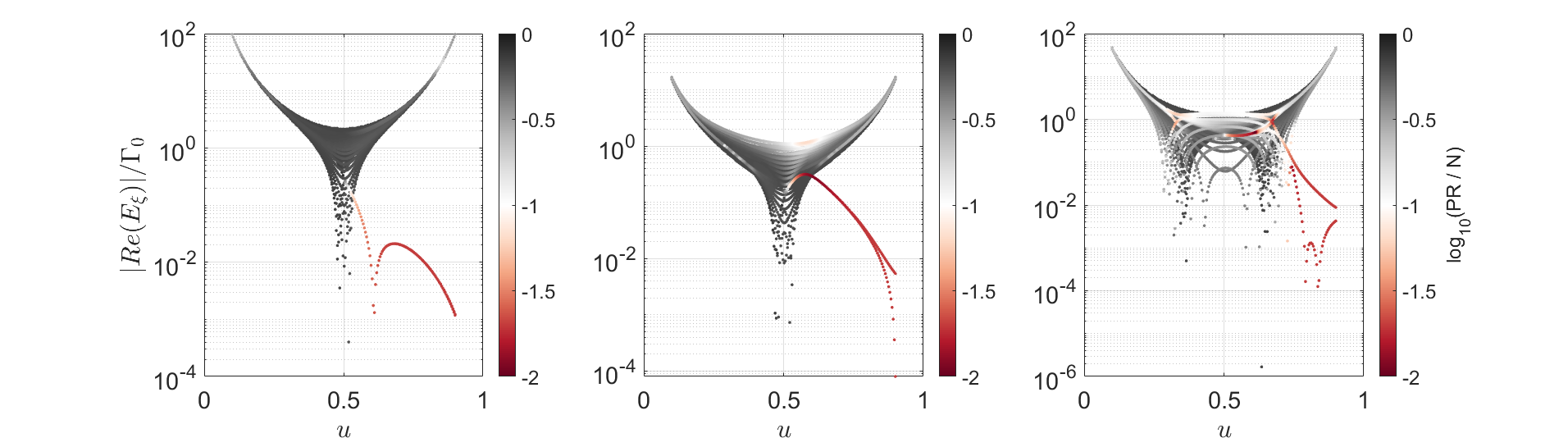} \put(6,28){\small\bfseries (a)}\put(36,28){\small\bfseries (b)}
    \put(65,28){\small\bfseries (c)}
    \end{overpic}
    \caption{Plots of absolute values of energies $|\Re(E)|/\Gamma_0$ vs $u$ in logarithmic scale. Different dipole orientations (a) $\theta = 0$, (b) $\theta = \cos^{-1}(1/\sqrt{3})-0.1$, (c) $\theta = \pi/2 $ are shown. The colorbars show participation ratios of the corresponding states in biorthonormal basis. Here, the chain consists of $N = 50$ unit cells with intercell spacings $a = 0.4 \lambda_0$.}
    \label{fig:reevsbovera}
\end{figure*}

For $\theta = 0$, two edge states with mid-gap energies and decay rates near $\Gamma=\Gamma_0$ appear for $u>0.5$ as deduced from bulk properties in Sec.~\ref{infinite}. Probability distributions of the edge states for same parameters but various $u$ values are plotted in Fig.~\ref{fig:NvsGamma}(b) where it can be seen that the localization at the boundaries strengthens as the ratio of intracell spacing to intercell spacing decreases. However, in the absence of a gap, the bulk properties discussed in Sec.~\ref{infinite} no longer predict the occurrence of edge states. In Fig.~\ref{fig:reevsbovera}, the energy shifts $\abs{\Re(E)/\Gamma_0}$ vs $u$ are plotted for various dipole orientations. Although the energies do not appear in $\pm E$ pairs due to the absence of chiral symmetry, the spectrum still consists of one negative and one positive energy band. Consequently, a gap at zero is indicated by a nonzero minimum of $\abs{\Re(E)/\Gamma_0}\neq 0$. The color encodes the participation ratio (PR) in the biorthonormal basis, and quantifies the degree of spatial localization of the eigenstates, with smaller (larger) PR corresponding to more localized (more extended) states. For dipole orientations shown in Fig.~\ref{fig:reevsbovera}(a,b), the bulk spectrum is gapped for all values of $u$, and mid-gap edge states appear for $u \, \textgreater \, 0.5$. By contrast, for $\theta=\pi/2$, shown in Fig.~\ref{fig:reevsbovera}(c), a gap is present only over a limited range of $u$ values. Hence, for some values of $u \neq 0.5$, even if we have non-trivial bulk invariant, robust edge states are not guaranteed, since the absence of a protecting bulk gap allows edge modes to hybridize with bulk states. This behavior depends on $\theta$ and $u$ in agreement with the band crossings observed in the bulk spectrum under periodic boundary conditions, as illustrated in Fig.~\ref{fig:modelandinfinitechainbands}(c).

\begin{figure*}
    \centering
    \includegraphics[width=\textwidth]{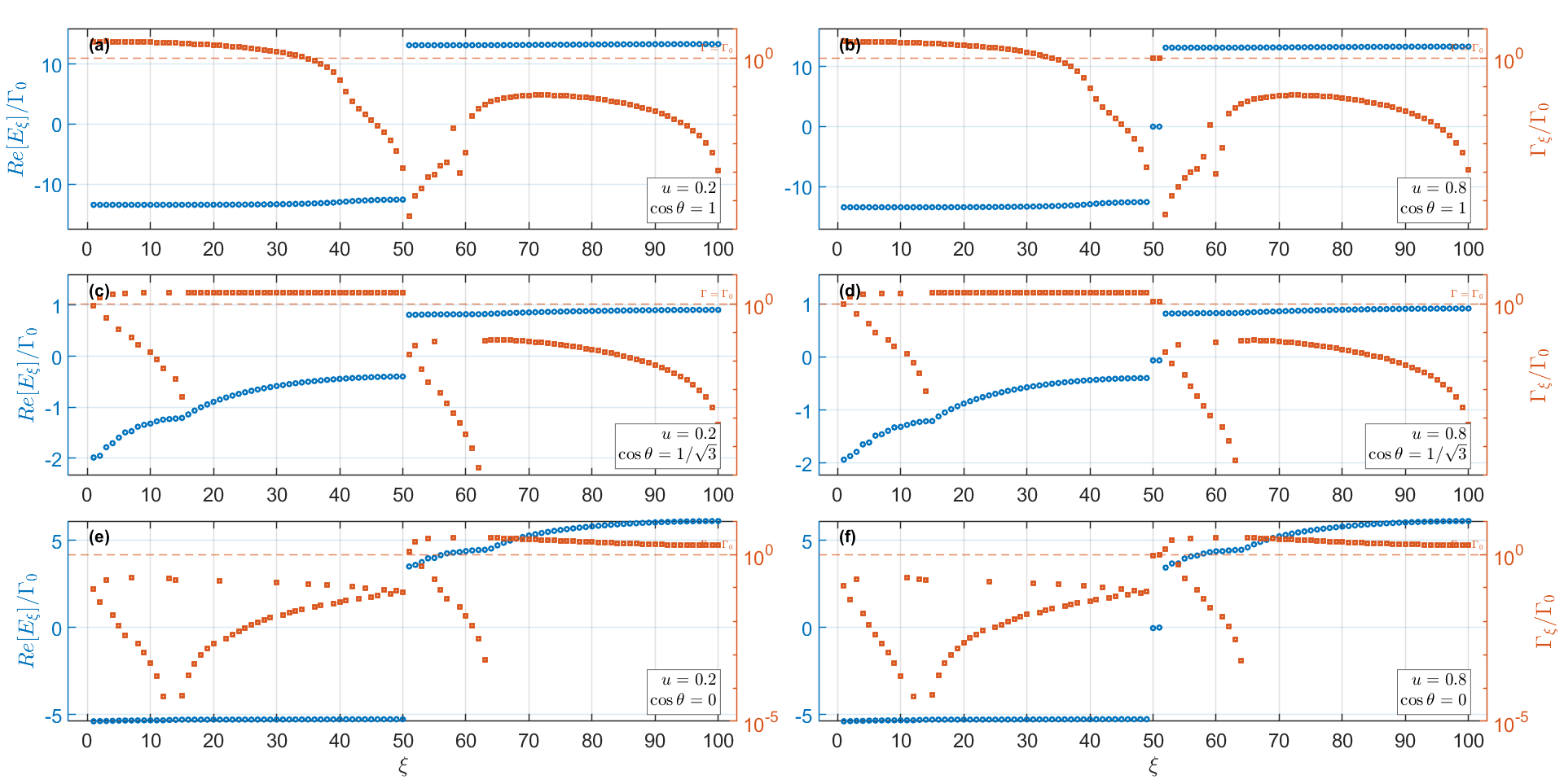}
    \caption{Real (blue-left axis) and imaginary (red-right axis) parts of the effective Hamiltonian's eigenvalues versus the eigenstate index $\xi$ for $a = 0.4 \lambda_0$ and $N = 50$ unit cells. In (a,c,e) $u = 0.2$ while in (b,d,f) $u = 0.8 $. The rows are different angular orientations (hence, different interatomic interaction scales). Figures (a,b) are for $\theta = 0$, (c,d) are for $\theta = \cos^{-1}(1/\sqrt{3})$, (e,f) are for $\theta = \pi/2$. Dashed red line shows  the value of $\Gamma_0$.}
    \label{fig:allangleenergies}
\end{figure*}
In Fig. \ref{fig:allangleenergies}, the ratios of the real (left axis, blue) and imaginary parts (right axis, red) of the eigenvalues to the individual atom decay rate; $\text{Re}[E_\xi]/\Gamma_0$ and $\text{Im}[E_\xi]/\Gamma_0$ are shown for $N=50$ unit cells, different dipole orientations (rows) and two different intracell spacings in gapped regime corresponding to the trivial ($u<0.5$ shown in left column) and non-trivial ($u>0.5$ shown in right column) topology. The real parts of the energies of bulk states form two bands separated by a gap characterized by $u$ and angle $\theta$. Compared to Fig. \ref{fig:allangleenergies}(c,d), where the dipole–dipole interaction contains only the $1/r$ term, the inclusion of $1/r^2$,$1/r^3$ terms result in a wider energy gap as shown in Fig. 6(a,b,e,f).

%% file: radiation.tex
\section{\label{Radiation}Radiation}
Until now, we mainly considered the dissipative dynamics of the chain with a collective excitation. We will now lean on the radiation pattern after the emission from different states. Until the excitation decays into the vacuum, the chain evolves continuously but non-unitarily under the effective Hamiltonian introduced in Eq.~\eqref{effectivehamiltonian}. Hence, the chain remains in an eigenstate of $H_{\rm eff}$ \cite{Orszag2016,Barnett1997,Daley2014}. After the chain emits a photon to the reservoir, it is in ground state $\ket{g} \equiv  \ket{g}^{\otimes N}$. In a similar manner, the field is in state $\ket{\Psi_{ph}(t)} = \sum_{q \nu} C_{q \nu}(t) \ket{1_{q\nu}}$ where the summation is over field modes ${\bf q}$ and polarizations $\nu$.  The time dependent state of the total system is
\begin{eqnarray}\label{timedepstate}
\begin{aligned}
    \ket{\Psi(t)} =&  \sum_{n\alpha} C_{n\alpha}(t) \sigma_+^{n\alpha} \ket{g} \otimes \ket{0} \\
    & + \sum_{q \nu} C_{q \nu}(t) \ket{g} \otimes \ket{1_{q\nu}}.
    \end{aligned}
\end{eqnarray}
Where the total Hamiltonian with rotating wave and dipole approximation in Schrödinger picture is
\begin{eqnarray}\label{radiationsecthamiltonian}
\begin{aligned}
    H &= \sum_{n \alpha} \frac{\hbar \omega_0}{2} \sigma_z^{n \alpha} + \sum_{q\nu} \hbar \omega_q  b_{q\nu}^{\dagger} b_{q\nu} \\&- i\sum_{\substack{q,\nu\\ n,\alpha}} g_{q\nu} (\sigma_{+}^{n\alpha}  b_{q\nu} e^{i {\bf q}\cdot {\bf r}_{n\alpha}}- \sigma_{-}^{n\alpha} b_{q\nu}^\dagger e^{-i {\bf q}\cdot {\bf r}_{n\alpha}} ).
    \end{aligned}
\end{eqnarray}
For identical atoms, with dipole moments parallel to each other, the coupling constant is, $g_{q\nu} = {\bf d} \cdot {{{\bf e}}}_{q\nu}\sqrt{\frac{\hbar \omega_q}{2 \epsilon_0 V}}$, where ${\bf d}$ is the dipole matrix element.  By solving the Schrödinger equation for Eq.~\eqref{radiationsecthamiltonian}, with the help of the initial condition indicating the absence of photons in the reservoir ($C_{q\nu}(0) = 0$), and the presence of a single excitation in the atomic system, we can obtain the photon amplitudes $C_{q\nu}(t)$ as 
\begin{equation}\label{cqnu}
C_{q\nu}(t) = \frac{1}{\hbar}\sum_{n\alpha} g_{q\nu} \,e^{-i {\bf q}\cdot {\bf r}_{n\alpha}} \int_0^t C_{n\alpha}(t') e^{-i\omega_{q}(t-t')} dt'.
\end{equation}  
Using the atomic amplitudes obtained in the previous sections, the photon amplitudes in Eq.\eqref{cqnu} can be evaluated explicitly. We will be interested in the dynamics of the singly-excited eigenstates of the effective Hamiltonian $H_{\rm eff}$.  For an infinite atomic chain, the atomic excitation takes the Bloch form $ C_{n\alpha}(t) = \frac{1}{\sqrt{N}} e^{-\frac{i}{\hbar}(\hbar \omega_0 + E_k)t} e^{ikna} u_{\alpha}(k)$ (see Sec.~\ref{infinite}). For a finite chain, the eigenvalues and eigenvectors of the effective Hamiltonian were obtained numerically in Sec.~\ref{finite} yielding $C_{n\alpha} (t) = e^{-\frac{i}{\hbar}(\hbar \omega_0 + E_\xi)t}v_{\xi; n\alpha} $ in Schrödinger picture.
\begin{figure*}[htbp]
    \centering

    \begin{minipage}[t]{0.32\textwidth}
        \begin{overpic}[width=\linewidth]{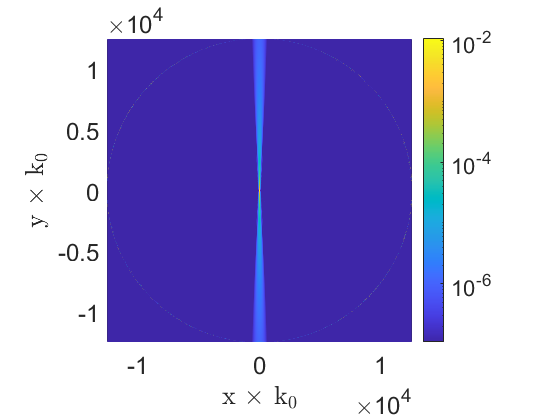}
            \put(0.5,70){\small\bfseries (a)}
        \end{overpic}
    \end{minipage}
    \hfill
    \begin{minipage}[t]{0.32\textwidth}
        \begin{overpic}[width=\linewidth]{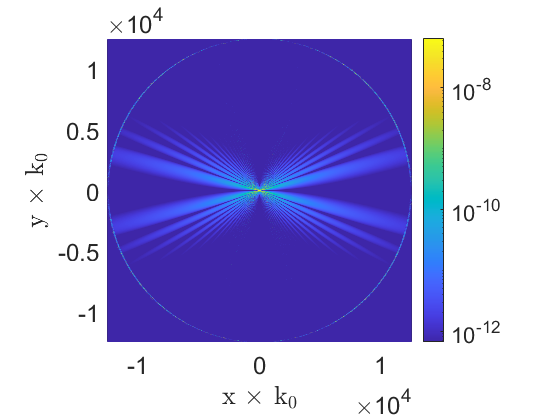}
            \put(0.7,70){\small\bfseries (b)}
        \end{overpic}
    \end{minipage}
    \hfill
    \begin{minipage}[t]{0.32\textwidth}
        \begin{overpic}[width=\linewidth]{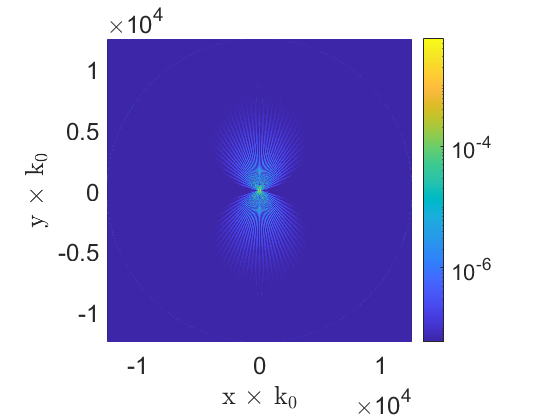}
            \put(0.7,70){\small\bfseries (c)}
        \end{overpic}
    \end{minipage}


    \begin{minipage}[t]{0.32\textwidth}
        \begin{overpic}[width=\linewidth]{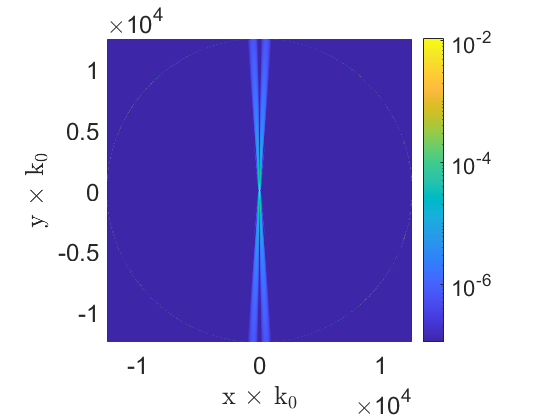}
            \put(0.7,70){\small\bfseries (d)}
        \end{overpic}
    \end{minipage}
    \hfill
    \begin{minipage}[t]{0.32\textwidth}
        \begin{overpic}[width=\linewidth]{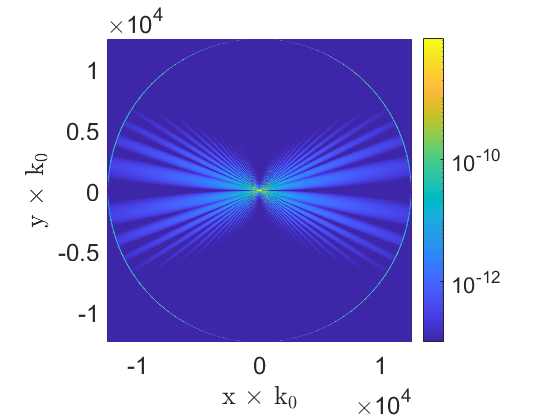}
            \put(0.7,70){\small\bfseries (e)}
        \end{overpic}
    \end{minipage}
    \hfill
    \begin{minipage}[t]{0.32\textwidth}
        \begin{overpic}[width=\linewidth]{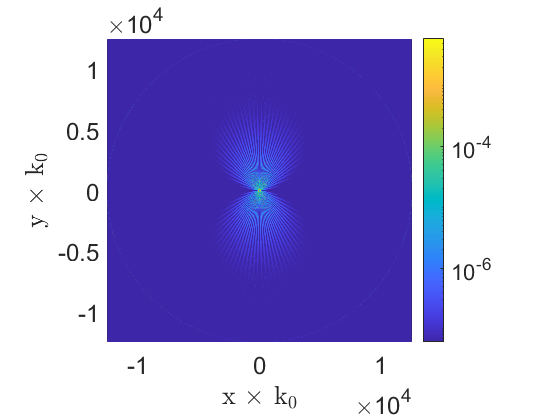}
            \put(0.7,70){\small\bfseries (f)}
        \end{overpic}
    \end{minipage}

    \caption{Far-field radiation patterns in the plane of the atomic array for $\theta=0$, $N=50$, $a=0.4\lambda_0$, $b=0.8a$, and $t=100 Na/c$. Panels (a--c) show emission from states with even unit-cell dipole-moment parity, while panels (d--f) show emission from states with odd parity. Panels (a,d) correspond to the two most superradiant states, panels (b,e) to the two most subradiant states, and panels (c,f) to the two edge states.}
    \label{fig:paraallelFarfieldpatterns}
\end{figure*}

\begin{figure*}[htbp]
    \centering

    \begin{minipage}[t]{0.32\textwidth}
        \begin{overpic}[width=\linewidth]{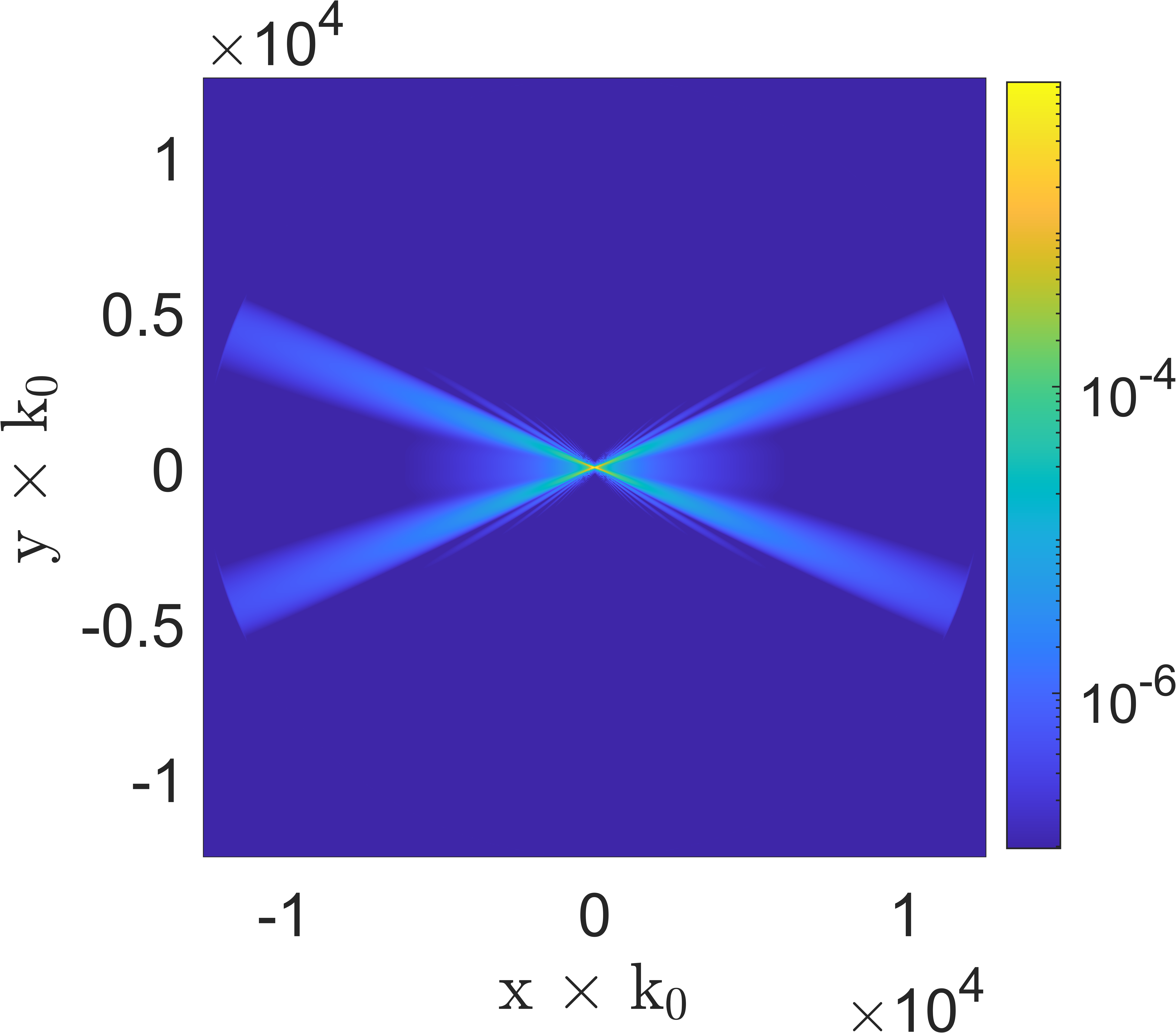}
            \put(0.5,70){\small\bfseries (a)}
        \end{overpic}
    \end{minipage}
    \hfill
    \begin{minipage}[t]{0.32\textwidth}
        \begin{overpic}[width=\linewidth]{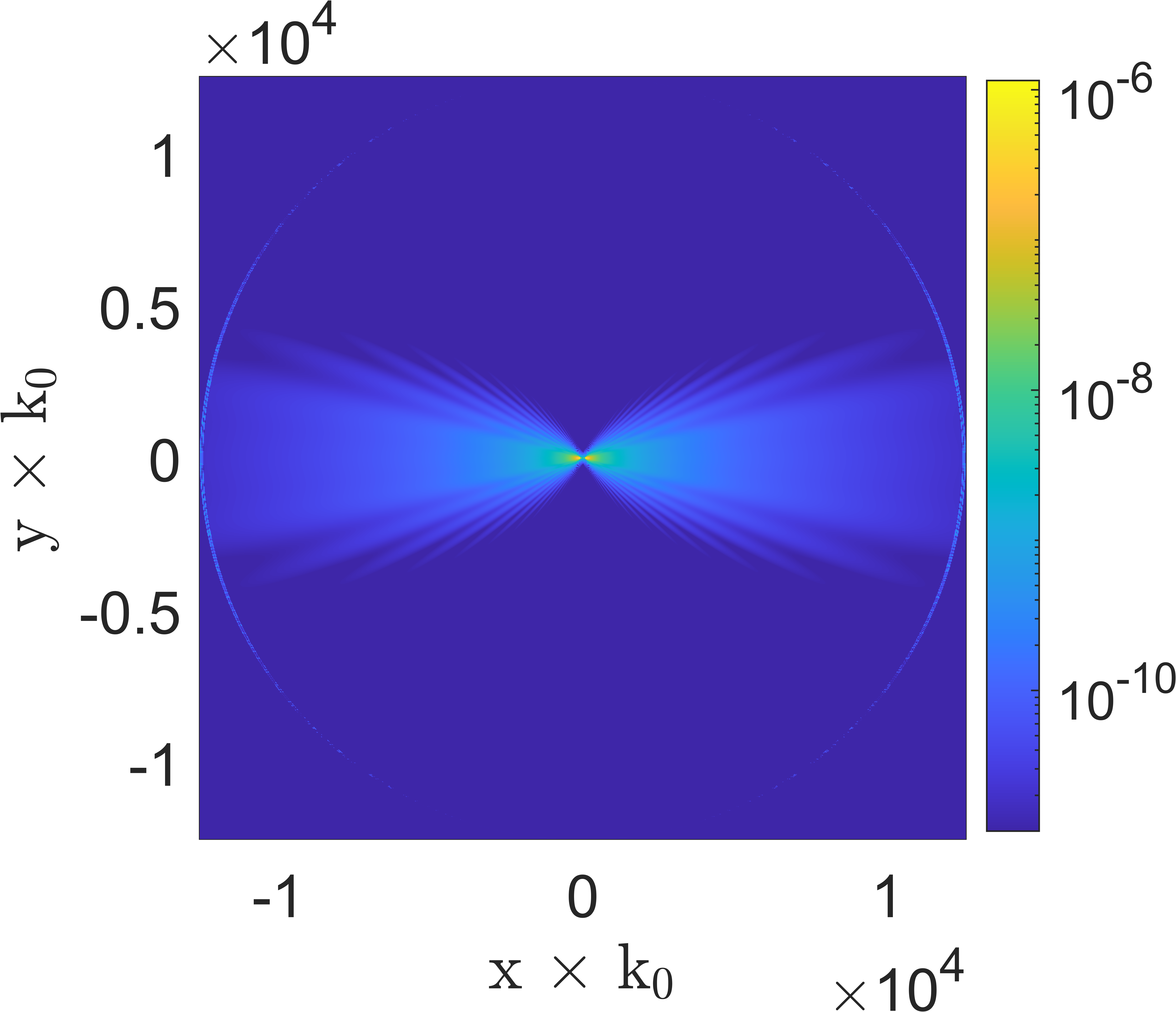}
            \put(0.7,70){\small\bfseries (b)}
        \end{overpic}
    \end{minipage}
    \hfill
    \begin{minipage}[t]{0.32\textwidth}
        \begin{overpic}[width=\linewidth]{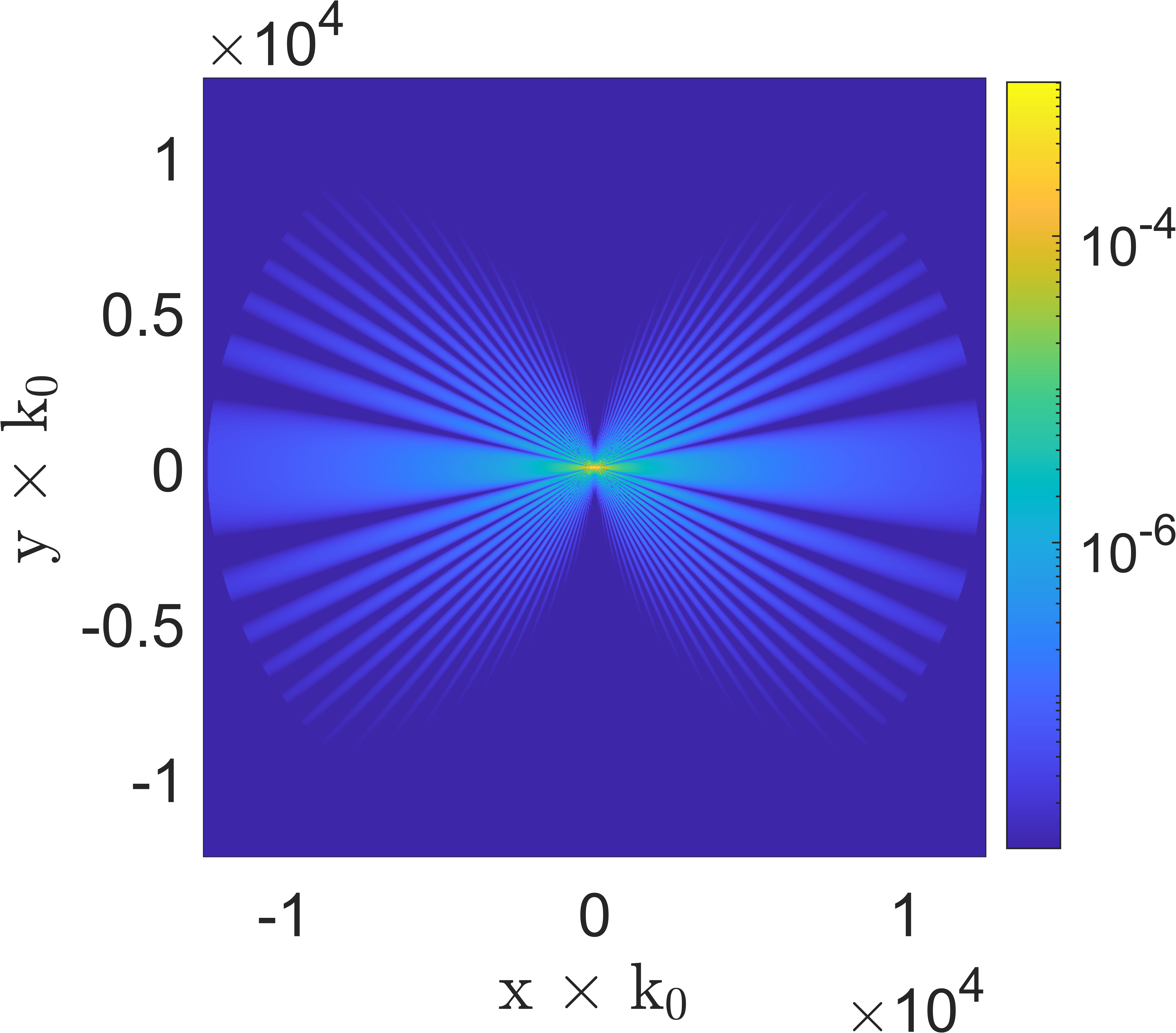}
            \put(0.7,70){\small\bfseries (c)}
        \end{overpic}
    \end{minipage}

    \begin{minipage}[t]{0.32\textwidth}
        \begin{overpic}[width=\linewidth]{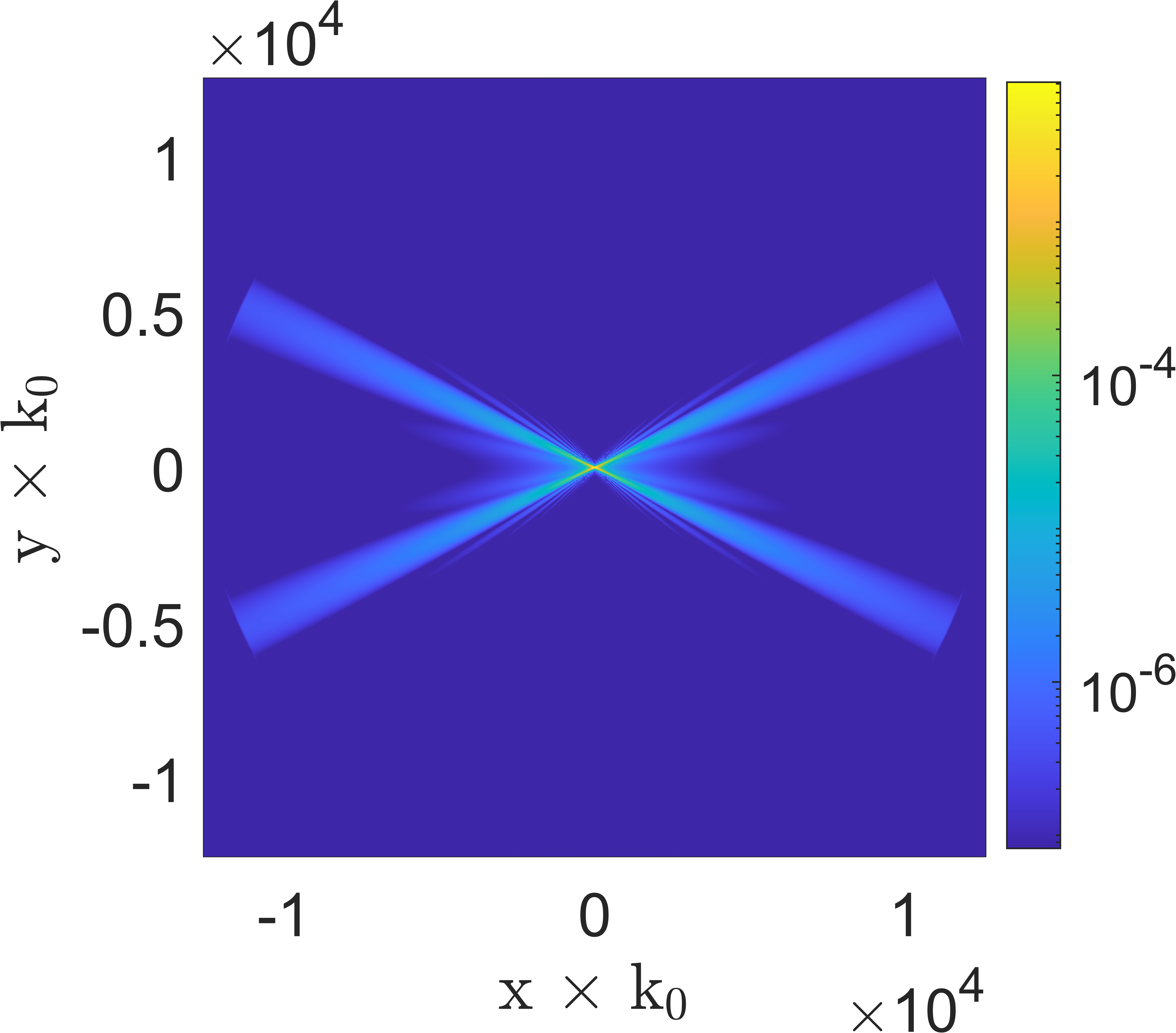}
            \put(0.7,70){\small\bfseries (d)}
        \end{overpic}
    \end{minipage}
    \hfill
    \begin{minipage}[t]{0.32\textwidth}
        \begin{overpic}[width=\linewidth]{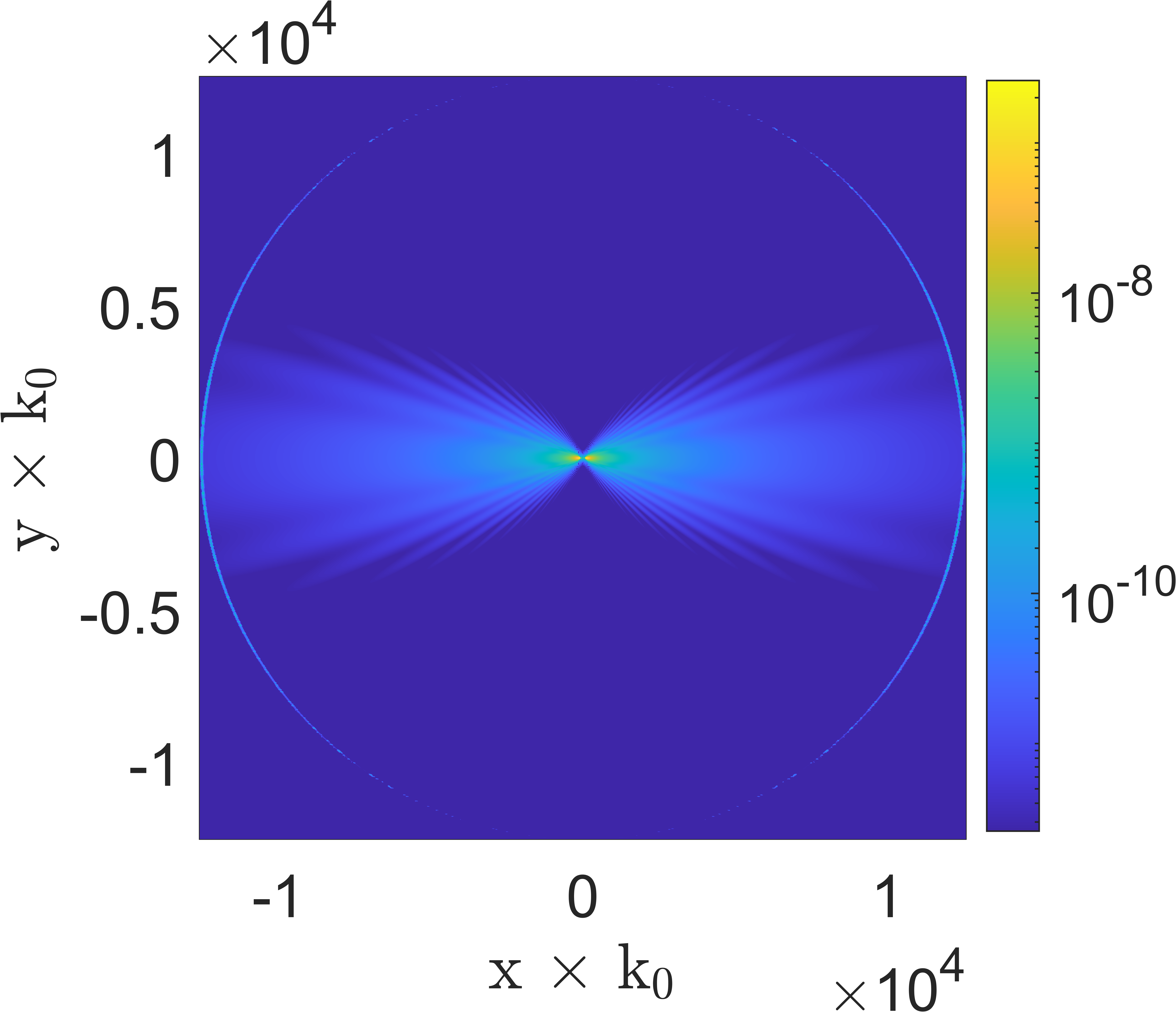}
            \put(0.7,70){\small\bfseries (e)}
        \end{overpic}
    \end{minipage}
    \hfill
    \begin{minipage}[t]{0.32\textwidth}
        \begin{overpic}[width=\linewidth]{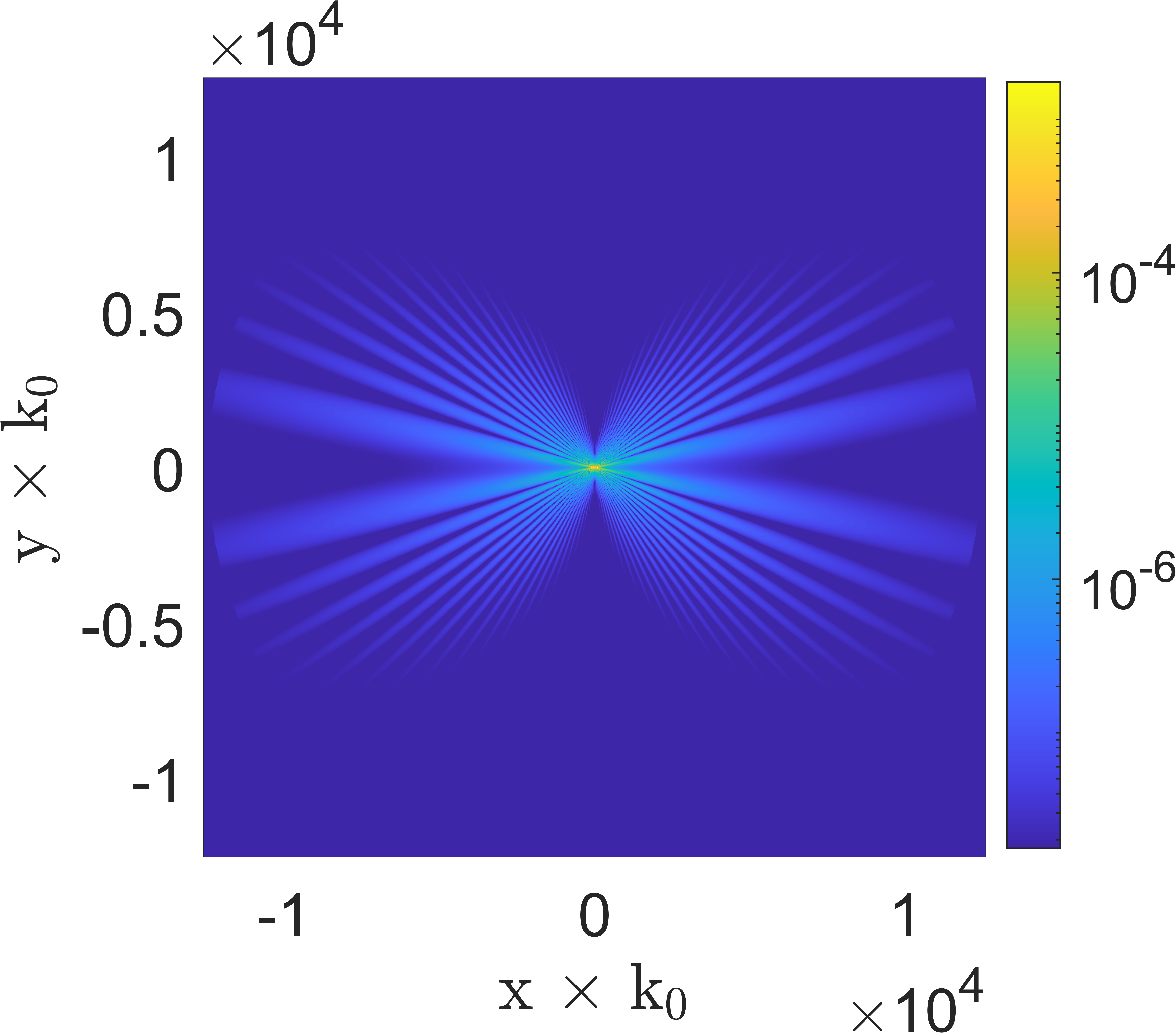}
            \put(0.7,70){\small\bfseries (f)}
        \end{overpic}
    \end{minipage}

    \caption{Far-field radiation patterns in the plane of the atomic array for $\theta=\pi/2$, $N=50$, $a=0.4\lambda_0$, $b=0.8a$, and $t=100 Na/c$. Panels (a--c) show emission from states with even unit-cell dipole-moment parity, while panels (d--f) show emission from states with odd parity. Panels (a,d) correspond to the two most superradiant states, panels (b,e) to the two most subradiant states, and panels (c,f) to the two edge states.}
    \label{fig:Perpendicularfarfieldpatterns}
\end{figure*}

After the decay from the atomic chain, the Poynting vector of the resulting radiation field can be determined by taking the expectation value over the photon state $\ket{\Psi_{ph}}$:
\begin{eqnarray}\label{poyntingvector}
\begin{aligned}
    \expval{{\bf S}}&= \frac{1}{\mu_0} \expval{:{{{\bf E}({\bf r})}} \times{{{\bf B}({\bf r})}}:}\\ &=\frac{1}{\mu_0} \expval{ \left({\bf E}^{(-)}({\bf r}) \times {{{\bf B}}}^{(+)}({\bf r})- {{{\bf B}}}^{(-)}({\bf r})\times {{{\bf E}}}^{(+)}({\bf r})\right)}.
\end{aligned}
\end{eqnarray}
Since ${{{\bf E}}}^{(+)} \times {{{\bf B}}}^{(+)}$ and ${{{\bf E}}}^{(-)} \times {{{\bf B}}}^{(-)}$ do not preserve particle number, their expectation values vanish. We can use the completeness, and obtain the following nonzero contributions.
 \begin{equation}\label{nonzeropoyntingterms}
     \begin{aligned}
       \expval{{\bf S}} = &\frac{1}{\mu_0} \Big{(}\bra{\Psi_{ph}(t)} {{{\bf E}}}^{(-)}({\bf r}) \ket{0} \times \bra{0} {{{\bf B}}}^{(+)}({\bf r}) \ket{\Psi_{ph}(t)}\\
      &- \bra{\Psi_{ph}(t)} {{{\bf B}}}^{(-)}({\bf r}) \ket{0} \times \bra{0} {{{\bf E}}}^{(+)}({\bf r}) \ket{\Psi_{ph}(t)}\Big{)} 
     \end{aligned}
 \end{equation}
Where the last two terms are complex conjugation of the rest. The term $\bra{0}{{{\bf E}}}^{(+)}({\bf r}) \ket{\Psi(t)}$ with negative frequency part of the electric field is
\begin{eqnarray}\label{epluspart}
\begin{aligned}
\bra{0} {{{{\bf E}}}} ^{(+)}({\bf r}) \ket{\Psi_{ph}(t)} = i \sum_{q\nu} \sqrt{\frac{\hbar\omega_q}{2\varepsilon_0 V} } \,{{\bf e}}_{q\nu}\, e^{i {\bf q}\cdot {\bf r}} C_{q\nu}(t)
\end{aligned}
\end{eqnarray}
with \eqref{cqnu} and $g_{q\nu}$ substituted, we obtain;
\begin{widetext}
\begin{eqnarray}
    \begin{aligned}
       \bra{0}{{{\bf E}}}^{(+)}({\bf r}) \ket{\Psi_{ph}(t)} =& \frac{i}{2\epsilon_0 V} \sum_{\substack{q\nu \\ n\alpha}} \omega_q ({\bf d} \cdot {{\bf e}}_{{\bf q}\nu}) {{\bf e}}_{{\bf q}\nu}\,e^{i {\bf q}\cdot ({\bf r}-{\bf r}_{n\alpha})} \int_0^t C_{n\alpha}(t') e^{-i\omega_{q}(t-t')} dt' \\
       =& \sum_{n\alpha}\int_0^t K_E({ {\bf r} - {\bf r}_{n\alpha},t;0,t'}) \, C_{n\alpha}(t') dt'
    \end{aligned}
\end{eqnarray}
\end{widetext}
where ${\bm{\mathscr{r}}}_{n\alpha}= {\bf r} - {\bf r}_{n\alpha}$, $\tau = t-t'$, and the Green function is;
\begin{eqnarray}
\begin{aligned}
   &K_E({\bm{\mathscr{r}}}_{n\alpha},t;0,t')  \\
   &= i \sum_{q\nu} \frac{\omega_{q}}{2 \varepsilon_0 V} ({\bf d} \cdot {{\bf e}}_{{\bf q}\nu}) {{\bf e}}_{{\bf q}\nu} e^{i ({\bf q} \cdot {\bm{\mathscr{r}}}_{n\alpha} - \omega_q \tau)}
\end{aligned}
\end{eqnarray}
The summation over the polarizations $\nu$ of the radiation can be taken beforehand. The only $\nu$ dependence is in the $({\bf d}\cdot{{\bf e}}_{q\nu}){{\bf e}}_{q\nu}$ term which is mutual for both infinite and finite chains:
\begin{eqnarray} \label{ddotetimesefinal}
    \sum_{\nu} ( {\bf d} \cdot  {{\bf e}}_{q\nu})  {{\bf e}}_{q\nu} =  {\bf d} - ( {\bf d} \cdot \hat{q}) \hat{q}.
\end{eqnarray}

By substituting Eq.~\eqref{ddotetimesefinal} and rewriting, Eq.~\eqref{epluspart} becomes;
\begin{eqnarray}
\begin{aligned}
     K_E({\bm{\mathscr{r}}}_{n\alpha},t;0,t') =&\\ \frac{ic}{2 \varepsilon_0 V}\sum_{{\bf q}}& \frac{1}{q} \left( {\bf d}\nabla^2 - \nabla({\bf d}\cdot\nabla) \right) e^{i ({\bf q} \cdot {\bm{\mathscr{r}}}_{n\alpha} - \omega_q \tau)}
\end{aligned}
\end{eqnarray}
In continuum limit $\sum_{ {\bf q}} \rightarrow \frac{V}{(2\pi)^3} \int d^3q$, we have;
\begin{eqnarray}
\begin{aligned}
&K_E({\bm{\mathscr{r}}}_{n\alpha},t;0,t')\\ & =\frac{ic}{2 (2\pi)^3 \varepsilon_0 }\left( {\bf d}\nabla^2 - \nabla({\bf d}\cdot\nabla) \right) \int \frac{1}{q} e^{i ({\bf q} \cdot {\bm{\mathscr{r}}}_{n\alpha} - \omega_q \tau)} d^3q\\
     &= \frac{i c}{2 (2\pi)^3 \varepsilon_0 }\left( {\bf d}\nabla^2 - \nabla({\bf d}\cdot\nabla) \right) \frac{2\pi^2}{i\mathscr{r}_{n\alpha} c} \delta( \tau - \mathscr{r}_{n\alpha}/c)
     \end{aligned}
\end{eqnarray}
Then, for Eq.~\eqref{epluspart} we have, 
\begin{eqnarray} \label{eplusgreensfunctionsubstituted}
\begin{aligned}
    &\bra{0}{{\bf E}}^{(+)} \ket{\Psi_{ph}(t)} = \int_0^t K_E({\bm{\mathscr{r}}}_{n\alpha},t;0,t') C_{n\alpha}(t') dt'\\
    &= \frac{1}{8\pi \varepsilon_0 } \sum_{n\alpha}  \left( {\bf d}\nabla^2 - \nabla({\bf d}\cdot\nabla) \right)  \left(\frac{ C_{n\alpha}(t-\mathscr{r}_{n\alpha}/c)}{\mathscr{r}_{n\alpha}} \right)
    \end{aligned}
\end{eqnarray}
Carrying out the differentiations and using the chain rule to express the derivatives in terms of the retarded time $t-\mathscr{r}/c$, we obtain the electric field contribution to the Poynting vector as follows:

\begin{widetext}
\begin{eqnarray}\label{eplusfinal}
\begin{aligned}
    \bra{0}{{\bf E}}^{(+)} \ket{\Psi_{ph}(t)} = \frac{1}{8\pi \varepsilon_0 } \sum_{n\alpha} & {\bf d} \left( \frac{\Ddot{C}_{n\alpha}}{c^2 \mathscr{r}_{n\alpha}} + \frac{\Dot{C}_{n\alpha}}{c \mathscr{r}_{n\alpha}^2}  + \frac{C_{n\alpha}}{\mathscr{r}_{n\alpha}^3}  \right)  &- \left( {\bf d} \cdot \hat{\mathscr{r}}_{n\alpha} \right) \hat{\mathscr{r}}_{n\alpha} \left( \frac{\Ddot{C}_{n\alpha}}{c^2 \mathscr{r}_{n\alpha}} +\frac{3 \, \Dot{C}_{n\alpha}}{c\mathscr{r}_{n\alpha}^2}  + \frac{3 \, C_{n\alpha}}{\mathscr{r}_{n\alpha}^3}  
       \right)
    \end{aligned}
\end{eqnarray}
\end{widetext}
Similarly, $\bra{\Psi_{ph}(t)} {{\bf B}}^{(-)} \ket{0} $ can be written as: 
\begin{eqnarray}\label{bminuspart}
\begin{aligned}
\bra{\Psi_{ph}(t)} {{\bf B}}^{(-)} \ket{0} = -i \sum_{q\nu} \sqrt{\frac{\hbar}{2\varepsilon_0 V\omega_q} } \,({\bf q} \times {{\bf e}}_{q\nu})\, e^{-i {\bf q}\cdot {\bf r}} C^*_{q\nu}
\end{aligned}
\end{eqnarray}
and substitution yields;
\begin{eqnarray}
    \begin{aligned}
       \bra{\Psi_{ph}(t)} {{\bf B}}^{(-)} \ket{0} = \sum_{n\alpha} \int_0^t K_B({{\bm{\mathscr{r}}}_{n\alpha},t;0,t'}) \, C^*_{n\alpha}(t') dt'
    \end{aligned}
\end{eqnarray}
When the summation over polarization $\nu$ is taken, we get:
\begin{eqnarray}
    \begin{aligned}
    K_B({\bm{\mathscr{r}}}_{n\alpha},t;0,t') =\frac{1}{2 \varepsilon_0 V} \sum_{q }  ({\bf d} \cross \nabla) e^{-i ({\bf q} \cdot {\bm{\mathscr{r}}}_{n\alpha} - \omega_q \tau)}
    \end{aligned}
\end{eqnarray}
Following the steps similar to those followed to obtain Eq.~\eqref{eplusfinal}, we take the continuum limit and obtain, we obtain the following expression for Eq.~\eqref{bminuspart}:
\begin{widetext}
\begin{eqnarray}\label{magneticfinal}
\begin{aligned}
     \bra{\Psi_{ph}(t)} {{\bf B}}^{(-)} \ket{0} = \frac{1}{8\pi \varepsilon_0 c} \sum_{ n,\alpha}  ({\bf d} \cross \hat{\mathscr{r}}_{n\alpha})  \left( \frac{1}{c^2 \mathscr{r}_{n\alpha}} \Ddot{C}^*_{n\alpha} + \frac{1}{c \, \mathscr{r}^2_{n\alpha}} \Dot{C}^*_{n\alpha} \right)
    \end{aligned}
\end{eqnarray}
\end{widetext}

 The Poynting vector Eq.~\eqref{nonzeropoyntingterms} is now obtained using Eqs. \eqref{eplusfinal} and \eqref{magneticfinal} as;
 \begin{eqnarray}\label{finalpoynting}
     \expval{{\bf S}} =\frac{1}{\mu_0} \left( \bra{0} {{{\bf E}}}^{(+)} \ket{\Psi_{ph}} \times  \bra{\Psi_{ph}} {{{\bf B}}}^{(-)} \ket{0} + h.c. \right)
 \end{eqnarray}
 
Eq.~\eqref{finalpoynting} gives the spatial and temporal distribution of the electromagnetic energy flux emitted by a given eigenstate of the non-Hermitian effective Hamiltonian $H_{\rm eff}$. For a finite chain with initial state $\ket{\Psi_\xi}$ given in Eq.~\eqref{finiteeigstate}, $C_{n\alpha}(t-\mathscr{r}_{n\alpha}/c) = v_{\xi; n\alpha}  e^{-i\,(\hbar \omega_0 + E_\xi) \, (t-\mathscr{r}_{n\alpha}/c)/\hbar}$ are found by diagonalizing $H_{\rm eff}$ numerically. We reveal that independent of the decay rate, some eigenstates exhibit a collective quadrupole-like radiation pattern due to vanishing \textit{unit cell dipole moment} while others exhibit a collective dipole radiation. More precisely, the unit-cell amplitudes $v_{\xi,n} = \sum_\alpha  v_{\xi; n\alpha} $ give vanishing or non-vanishing dipole moments depending on whether the amplitudes exhibit even or odd parity along the chain. These amplitudes for the considered states can be found in Appendix \ref{sec:nearfield}. Eq.~\eqref{finalpoynting} is plotted for six different $\ket{\Psi_\xi}$ in radiation zone for dipoles moments along the chain in Fig.\ref{fig:paraallelFarfieldpatterns} and perpendicular to the chain in Fig.\ref{fig:Perpendicularfarfieldpatterns}. In order to see the dependence of radiation pattern on the parity of $v_{\xi,n}$, pairs of states with similar decay rates but different dipole moment parity along the chain are shown explicitly. More specifically, Fig.~\ref{fig:paraallelFarfieldpatterns}(a-c) and Fig.~\ref{fig:Perpendicularfarfieldpatterns}(a-c) show radiation from states with even parity and Fig.~\ref{fig:paraallelFarfieldpatterns}(d-f) and Fig.~\ref{fig:Perpendicularfarfieldpatterns}(d-f) correspond to states with odd parity. From Fig.~\ref{fig:paraallelFarfieldpatterns} we see that, for $\theta = 0$, states whose unit-cell amplitudes have even parity exhibit constructive radiation in the direction perpendicular to the midpoint of the chain, whereas odd-parity states display destructive interference, leading to a vanishing field along this direction. A similar parity dependence appears for $\theta=\pi/2$ as illustrated in Fig.~\ref{fig:Perpendicularfarfieldpatterns}, but in this case the distinction is manifested primarily along the chain direction.

A more pronounced difference in the radiation patterns emerges when comparing superradiant, subradiant, and edge states, which exhibit distinct decay rates and localization properties. For $\theta=0$, the superradiant states exhibit a narrow intensity distribution that is maximized along the direction perpendicular to the chain, as shown in Fig.~\ref{fig:paraallelFarfieldpatterns}(a,d), reflecting the constructive interference of emitters across the bulk, leading to strongly directional radiation. For $\theta = \pi/2$, the direction perpendicular to the chain is the direction of the dipole moments. Hence, we do not observe an emission along this direction in Fig.~\ref{fig:Perpendicularfarfieldpatterns}(a,d), but instead of an intensity distribution similar to dipole radiation pattern, the peak positions lean towards the perpendicular direction. In contrast to the superradiant case, the emission from subradiant states is significantly suppressed within the bulk of the finite chain and occurs predominantly at the ends, as seen in Fig.~\ref{fig:paraallelFarfieldpatterns}(b,e) and Fig.~\ref{fig:Perpendicularfarfieldpatterns}(b,e), indicating destructive interference in the interior, with radiation primarily leaking out from the boundaries. This suppression is consistent with the unit-cell dipole moment profiles discussed in Appendix~\ref{sec:nearfield}, which suggest dominant large wave vector Fourier components that are weakly coupled to propagating radiation modes. This interpretation is more clearly supported by the near-field emission patterns shown in Appendix \ref{sec:nearfield}, since the finite chain is not spatially resolved in the far-field radiation plots. Lastly, the emission patterns associated with the edge states are shown in Fig.~\ref{fig:paraallelFarfieldpatterns}(c,f) and Fig.~\ref{fig:Perpendicularfarfieldpatterns}(c,f). These states are localized at both ends of the chain and exhibit radiation patterns similar to a pair of radiating dipoles, each centered at one edge with corresponding alignments of Fig.~\ref{fig:paraallelFarfieldpatterns} and Fig.~\ref{fig:Perpendicularfarfieldpatterns}.

%% file: conclusion.tex
\section{Conclusion}

In this work, we have analyzed the topological and collective radiative properties of a one-dimensional diatomic chain of identical quantum emitters coupled to the electromagnetic vacuum. The system realizes an extended, non-Hermitian Su–Schrieffer–Heeger–type model with all-to-all interactions and collective dissipation, described by an effective non-Hermitian Hamiltonian in the single-excitation manifold. 

For the infinite chain, we examined the complex band structure and identified regimes of subradiance associated with modes lying outside the light cone. We showed that, despite the absence of strict chiral symmetry due to long-range intra-sublattice couplings, inversion symmetry ensures a quantized complex Berry phase, allowing for a well-defined topological classification in gapped regions. We further identified parameter regimes where real-part band crossings occur, limiting the applicability of bulk–boundary correspondence.

For finite chains, exact diagonalization revealed superradiant, subradiant, and topological edge states with distinct decay rates and spatial localization properties. Subradiant states exhibit strongly suppressed decay with system-size-dependent scaling, while edge states emerge in the topologically nontrivial phase when a bulk gap is present. We showed that the existence and robustness of these edge states depend sensitively on the spectral gap.

Finally, we analyzed the far-field radiation patterns associated with different classes of eigenstates. The results demonstrate that collective radiation signatures reflect both the spatial structure and decay properties of these eigenstates, with clear distinctions between superradiant, subradiant, and edge modes, as well as unit cell dipole moment amplitude parity dependent emission characteristics.

%% file: lerchtranscendent.tex
\section{Lerch Transcendent Forms and Convergence} \label{lerchtranscendent}
The infinite sums over $\ell = n-m$ in Eqs. \eqref{eq:openforms} can be written in terms of the Lerch transcendent: 

\begin{equation}
    \Phi(z,s,\alpha) = \sum_{n=0}^{\infty} \frac{z^n}{(n+\alpha)^s}
\end{equation}
whose specializations recover the Hurwitz zeta ($ \zeta(s,\alpha) = \Phi(1,s,\alpha)$) and the polylogarithm ($Li_s(z) = z\Phi(z,s,1)$). Before presenting the rewritings open-form expressions, we introduce three auxiliary functions,
$g(n)$ and $f(n)$, defined in terms of polylogarithms and Lerch transcendents, for compactness:

\begin{equation}\label{gn}
    g(n)= \frac{1}{a^n} \left(Li_n(e^{-ia(k-k_0) - \varepsilon}) + 
   Li_n( e^{ia(k+k_0) - \varepsilon})\right)
\end{equation}
\begin{equation}\label{fab}
\begin{aligned}
        f_{AB}(s) = &\frac{e^{iua}}{a^s}\Phi(e^{ia(k + k_0) - \varepsilon},s,u)\\+& \frac{e^{-iua}}{a^s} \left(\Phi(e^{-ia(k - k_0) - \varepsilon},s, -u) - \frac{1}{u^s}\right)
\end{aligned}
\end{equation}
\begin{equation}\label{fba}
\begin{aligned}
    f_{BA}(s) =& \frac{e^{-iua}}{a^s}\left(\Phi(e^{ia(k + k_0) - \varepsilon},s, -u) - \frac{1}{u^s} \right) \\ + &  \frac{e^{iua}}{a^s} \Phi(
      e^{-ia(k - k_0) - \varepsilon}, s, u) 
\end{aligned}
\end{equation}

Here, $u=b/a$, and $e^{-\varepsilon \ell}$ serves as a convergence factor. For $\varepsilon=0$, the long-range interaction terms $f_{AB}(1)$ and $f_{BA}(1)$ reduce to harmonic series at $k=k_0$, which diverge, 
while the auxiliary functions converge otherwise. With the definitions in eqs. \eqref{gn}--\eqref{fba}, the open-form expressions from
Eq.~\eqref{eq:openforms} can now be compactly written in terms of these as follows:
\begin{widetext}
\begin{equation}
\begin{aligned}
G_{AA}(k) = \frac{3\Gamma_0}{4} \left[- c_1 g(1) + c_2 \left(-i g(2) +g(3)\right) \right]-\frac{i \Gamma_0}{2}, 
\end{aligned}
\label{eq:Gaa_full}
\end{equation}

\begin{equation}
\begin{aligned}
    G_{AB}(k) = \frac{3\Gamma_0}{4} \left[ - c_1 f_{AB}(1) + c_2 (-i f_{AB}(2) + f_{AB}(3))\right],
\end{aligned}
\end{equation}

\begin{equation}
    \begin{aligned}
        G_{BA}(k) = \frac{3\Gamma_0}{4} \left[ -\left[ 1- \cos^2\theta \right] f_{BA}(1) + \left[ 1- 3\cos^2\theta \right] (-i f_{BA}(2)+f_{BA}(3)) \right]
    \end{aligned}
\end{equation}
\end{widetext}



%% file: nearfield.tex
\section{\label{sec:nearfield} Near Field Patterns and Unit Cell Dipole Moments}

Below, Fig.\ref{fig:parallelnearfieldpatterns} and Fig.\ref{fig:perpendicularnearfieldpatterns} show near zone patterns of emission from different eigenstates of $H_{\rm eff}$ for parallel and perpendicular dipoles respectively. For edge states, $t=0.8 Na/c$ while for other states, $t=0.5 Na/c$. Since the considered times are much shorter than the lifetime of the corresponding collective excitation, the excitation remains predominantly stored in the atomic chain. Consequently, the near-zone field contains very large contributions close to the atomic positions, which dominate the color scale and hide the weaker field pattern in the surrounding region. To make the surrounding radiation pattern visible, values exceeding a chosen upper threshold were masked in the plots. This cutoff is used only for visualization and does not alter the calculated field distribution.

\begin{figure*}[!htbp]
    \centering

    \begin{minipage}[t]{0.325\textwidth}
        \begin{overpic}[width=\linewidth]{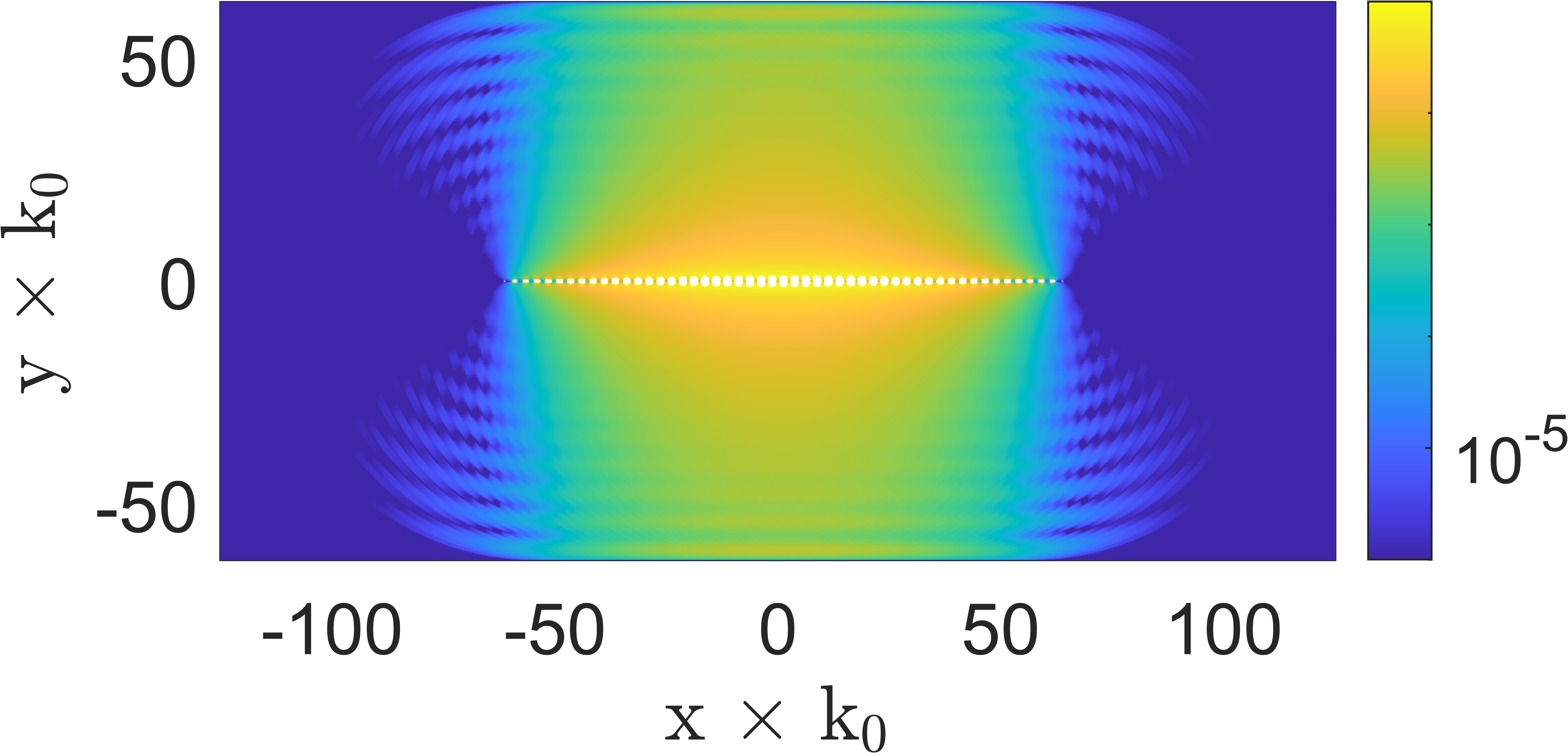}
            \put(0.0001,50){\small\bfseries a)}
        \end{overpic}
    \end{minipage}
    \hfill
    \begin{minipage}[t]{0.325\textwidth}
        \begin{overpic}[width=\linewidth]{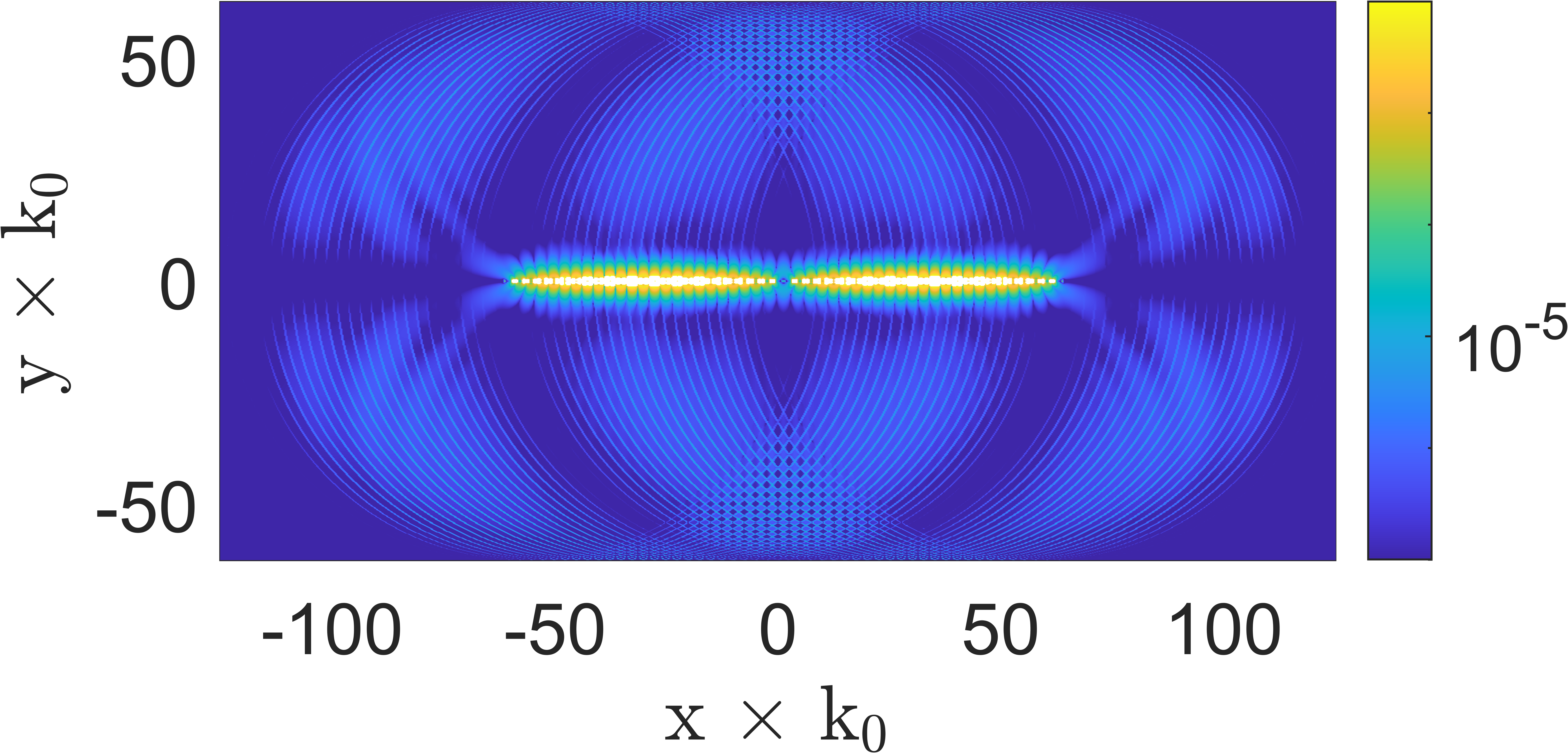}
            \put(0.00001,50){\small\bfseries b)}
        \end{overpic}
    \end{minipage}
    \hfill
    \begin{minipage}[t]{0.325\textwidth}
        \begin{overpic}[width=\linewidth]{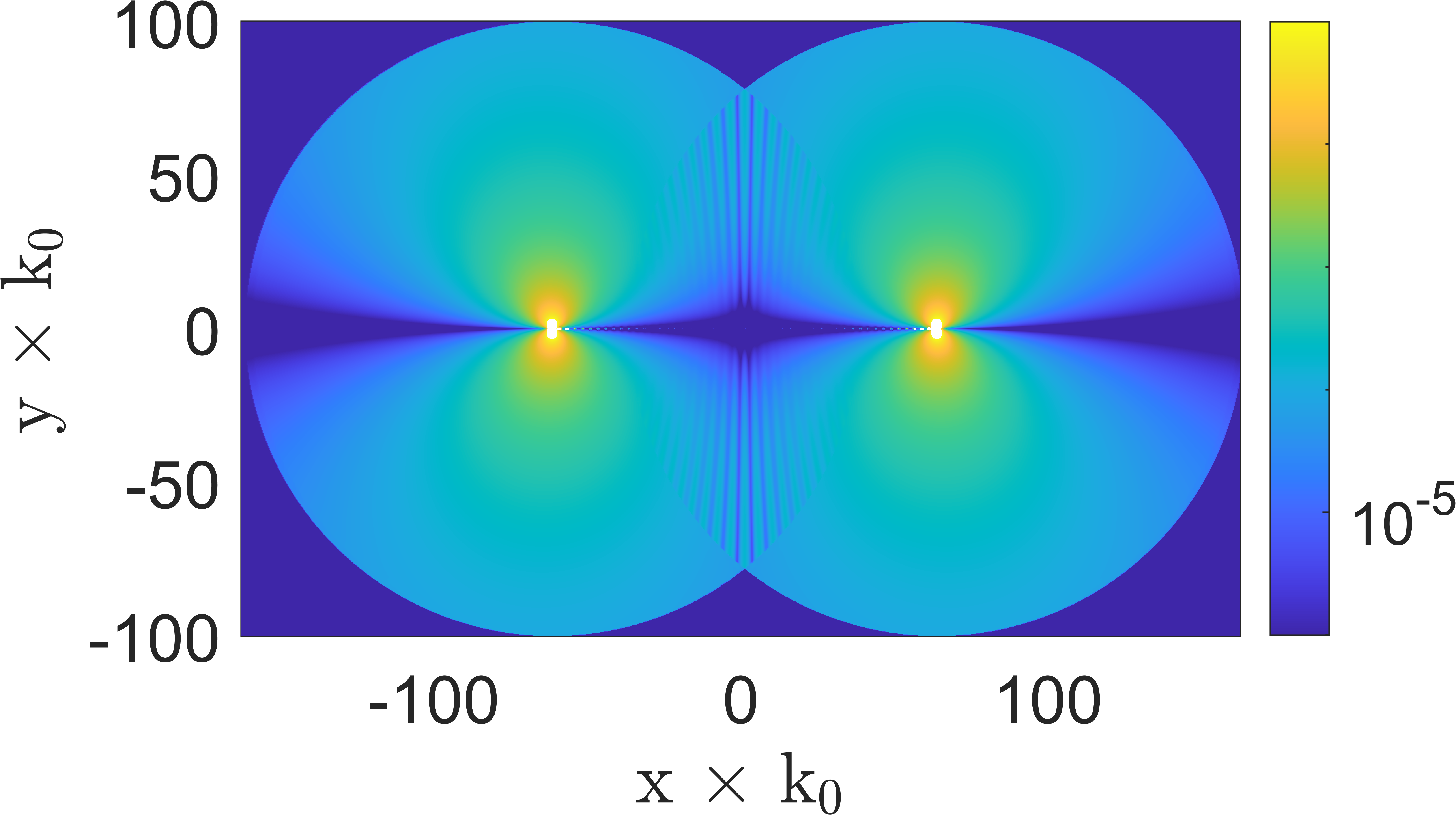}
            \put(0.0000001,50){\small\bfseries c)}
        \end{overpic}
    \end{minipage}


    \begin{minipage}[t]{0.325\textwidth}
        \begin{overpic}[width=\linewidth]{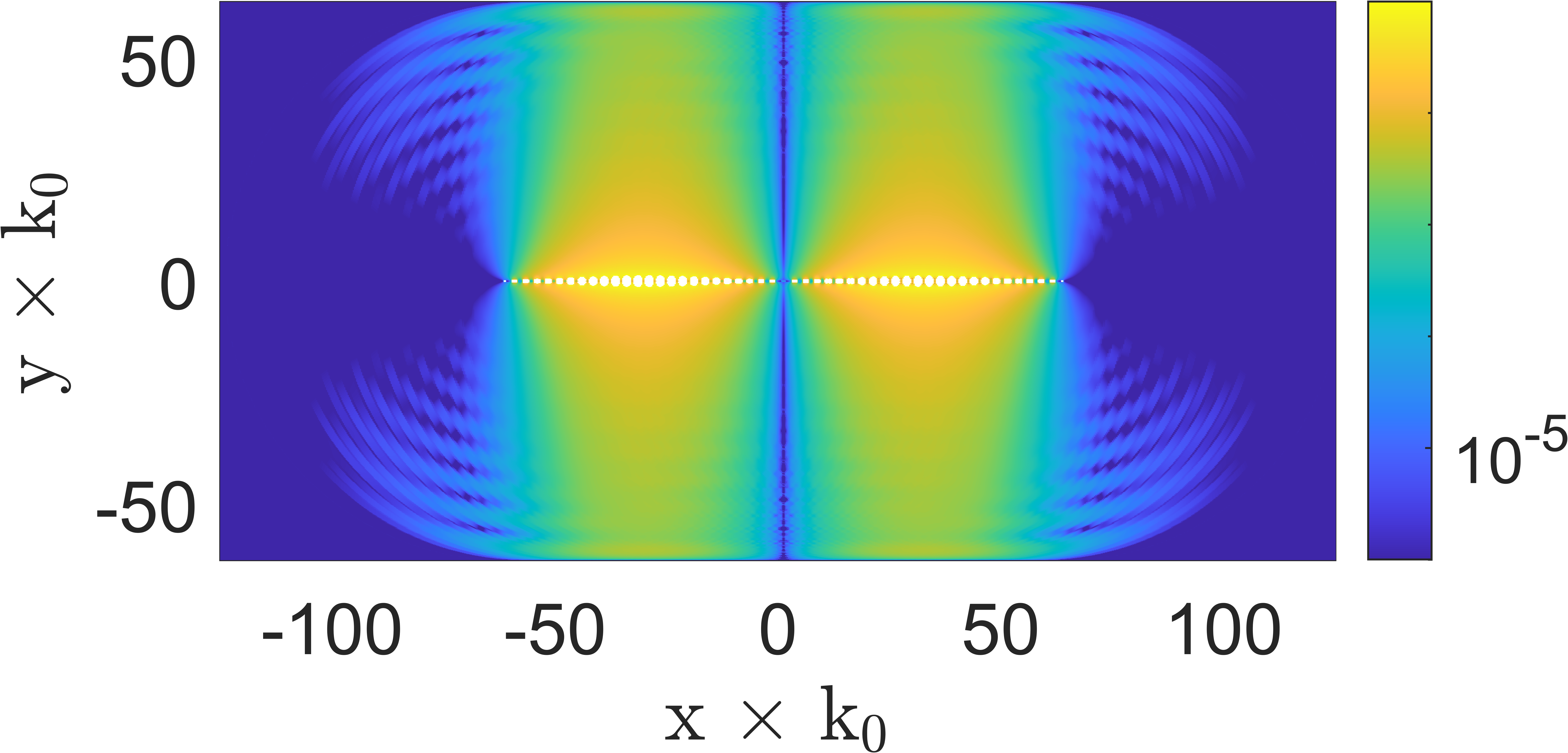}
            \put(0.0000001,50){\small\bfseries d)}
        \end{overpic}
    \end{minipage}
    \hfill
    \begin{minipage}[t]{0.325\textwidth}
        \begin{overpic}[width=\linewidth]{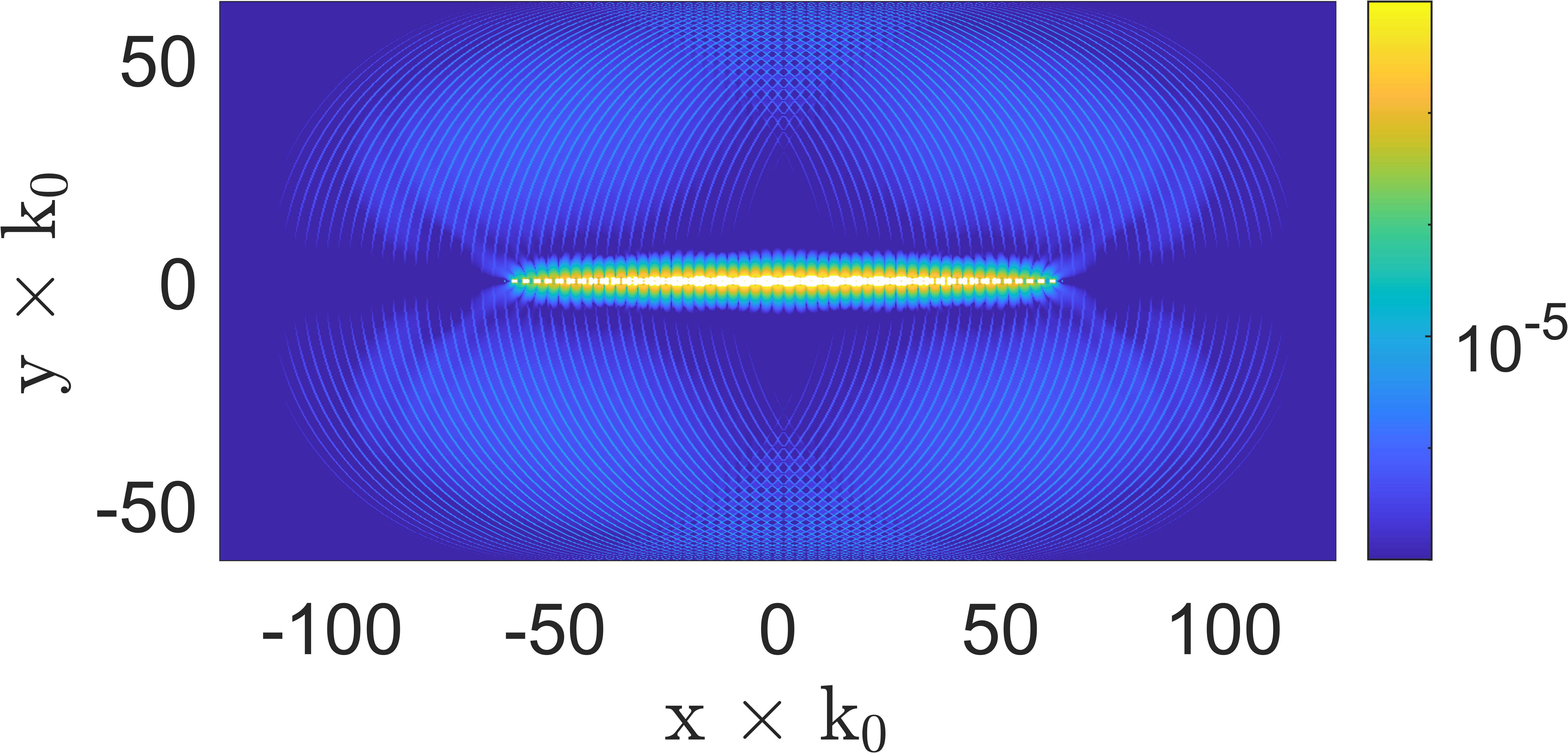}
            \put(0.0000001,50){\small\bfseries e)}
        \end{overpic}
    \end{minipage}
    \hfill
    \begin{minipage}[t]{0.325\textwidth}
        \begin{overpic}[width=\linewidth]{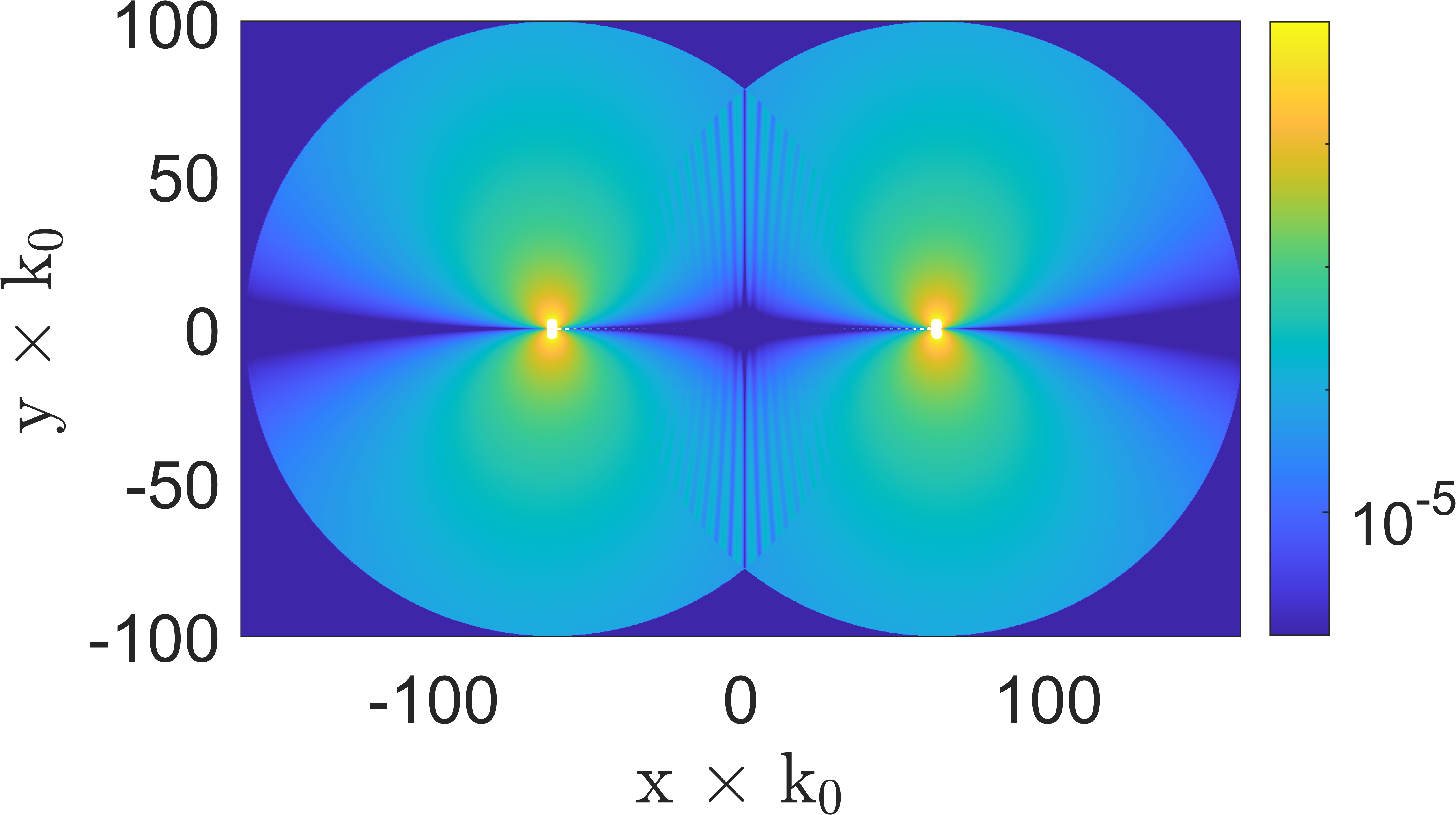}
            \put(0.000001,50){\small\bfseries f)}
        \end{overpic}
    \end{minipage}

    \caption{The near zone figures in the plane that atomic array lies; for $\theta=0$, $a=0.4 \lambda_0$, $u=0.8$, $t= 0.8 Na/c$ for the edge states and $t= 0.5 Na/c$ for the rest. The chain consists of $N=50$ unit cells. Panels (a--c) show emission from states with even unit-cell dipole-moment parity, while panels (d--f) show emission from states with odd parity. Panels (a,d) correspond to the two most superradiant states, panels (b,e) to the two most subradiant states, and panels (c,f) to the two edge states. Values above the upper color-scale threshold are masked to reveal the weaker near-zone spatial structure away from the chain. The threshold is chosen separately for each panel to improve visibility.}
    \label{fig:parallelnearfieldpatterns}

\end{figure*}
\begin{figure*}[!htbp]
    \centering

    \begin{minipage}[t]{0.325\textwidth}
        \begin{overpic}[width=\linewidth]{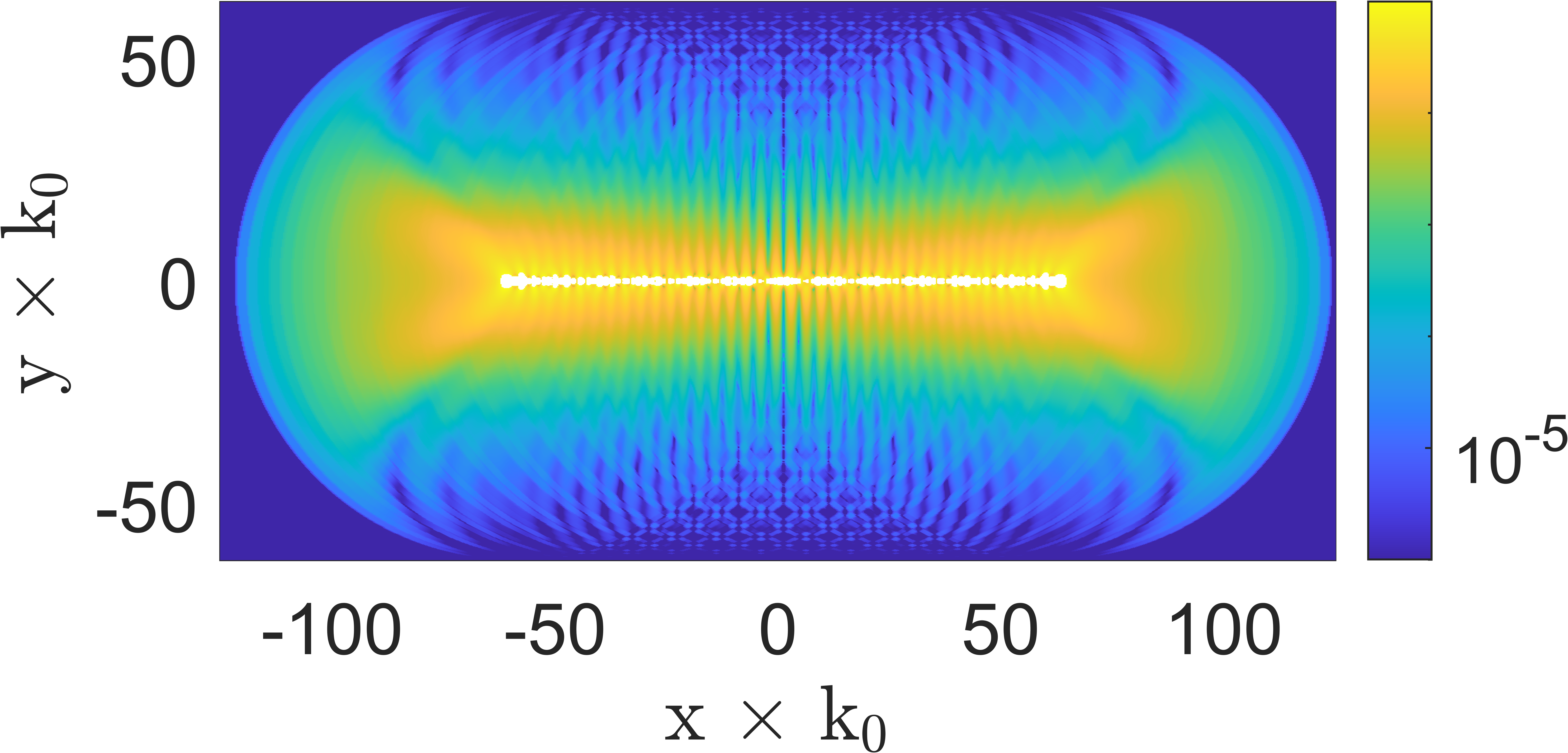}
            \put(0.0001,50){\small\bfseries a)}
        \end{overpic}
    \end{minipage}
    \hfill
    \begin{minipage}[t]{0.325\textwidth}
        \begin{overpic}[width=\linewidth]{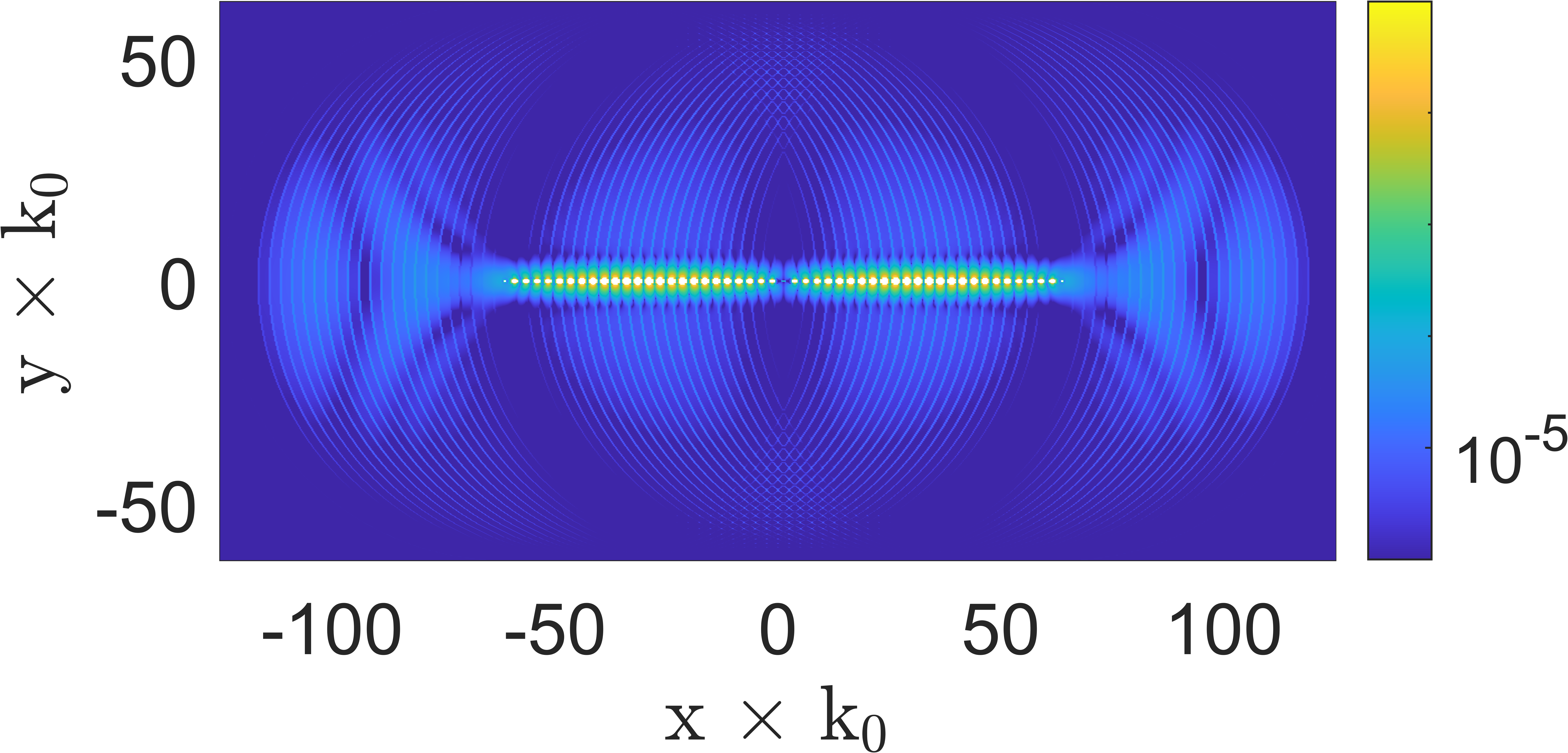}
            \put(0.00001,50){\small\bfseries b)}
        \end{overpic}
    \end{minipage}
    \hfill
    \begin{minipage}[t]{0.325\textwidth}
        \begin{overpic}[width=\linewidth]{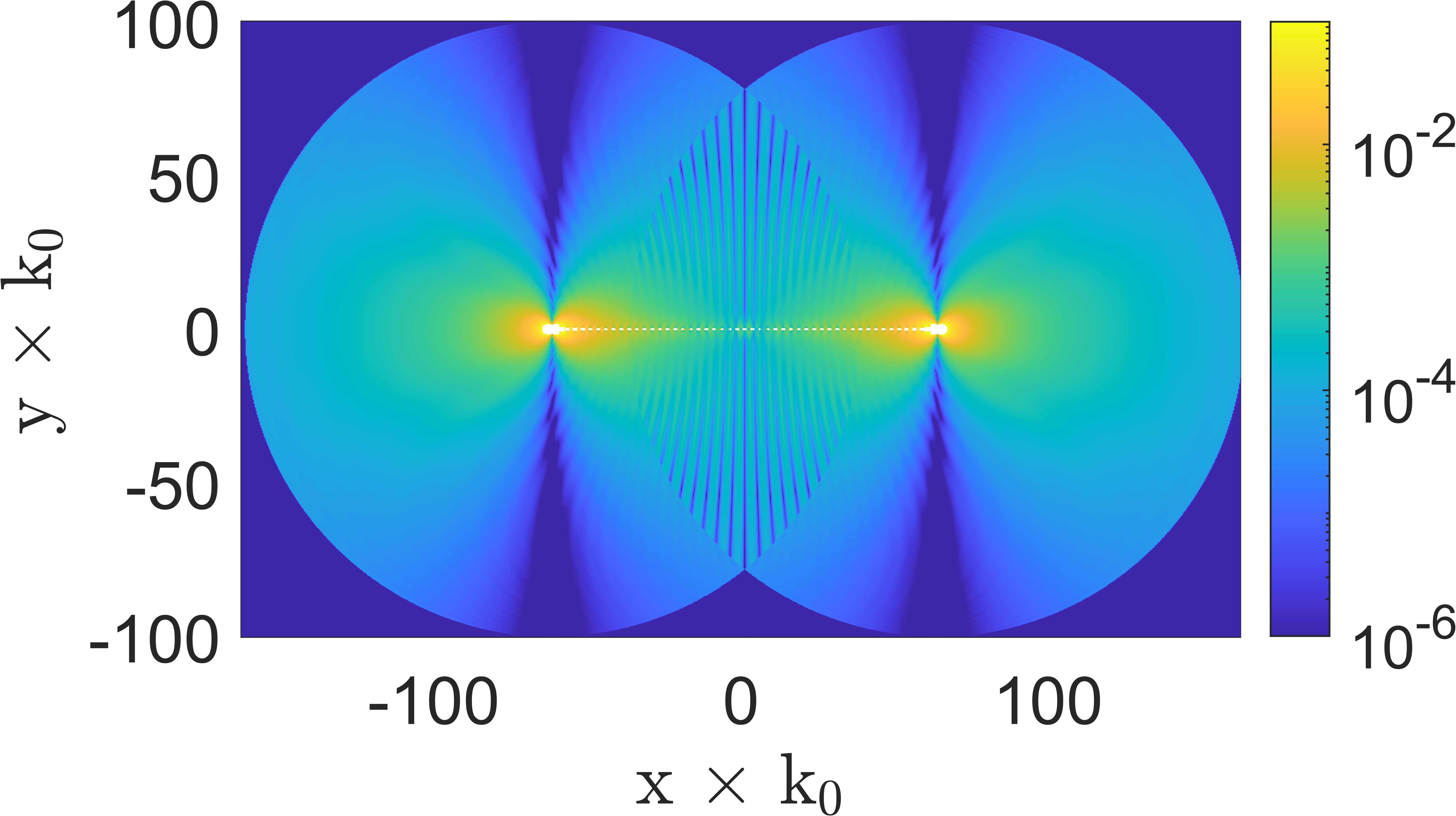}
            \put(0.0000001,50){\small\bfseries c)}
        \end{overpic}
    \end{minipage}


    \begin{minipage}[t]{0.325\textwidth}
        \begin{overpic}[width=\linewidth]{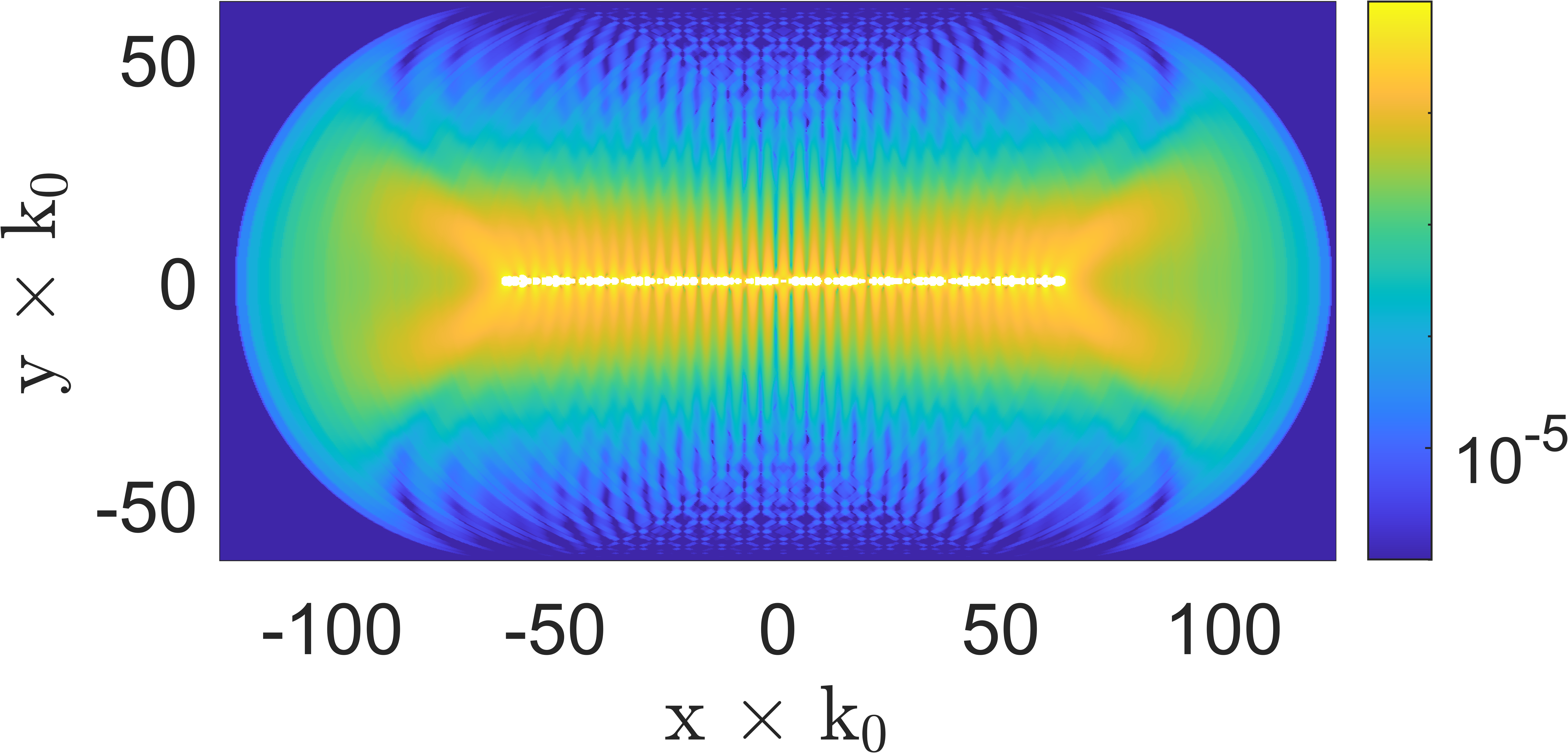}
            \put(0.0000001,50){\small\bfseries d)}
        \end{overpic}
    \end{minipage}
    \hfill
    \begin{minipage}[t]{0.325\textwidth}
        \begin{overpic}[width=\linewidth]{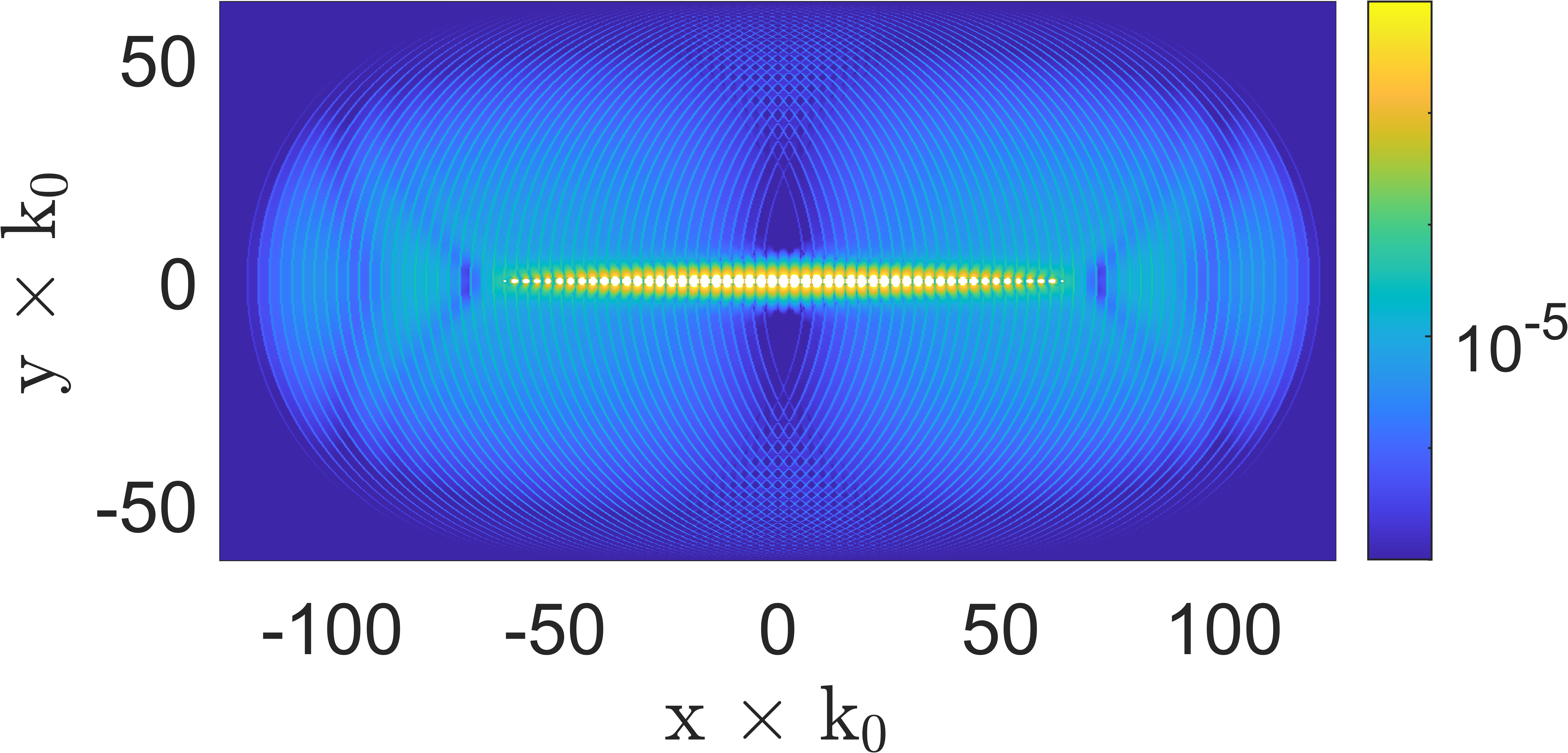}
            \put(0.0000001,50){\small\bfseries e)}
        \end{overpic}
    \end{minipage}
    \hfill
    \begin{minipage}[t]{0.325\textwidth}
        \begin{overpic}[width=\linewidth]{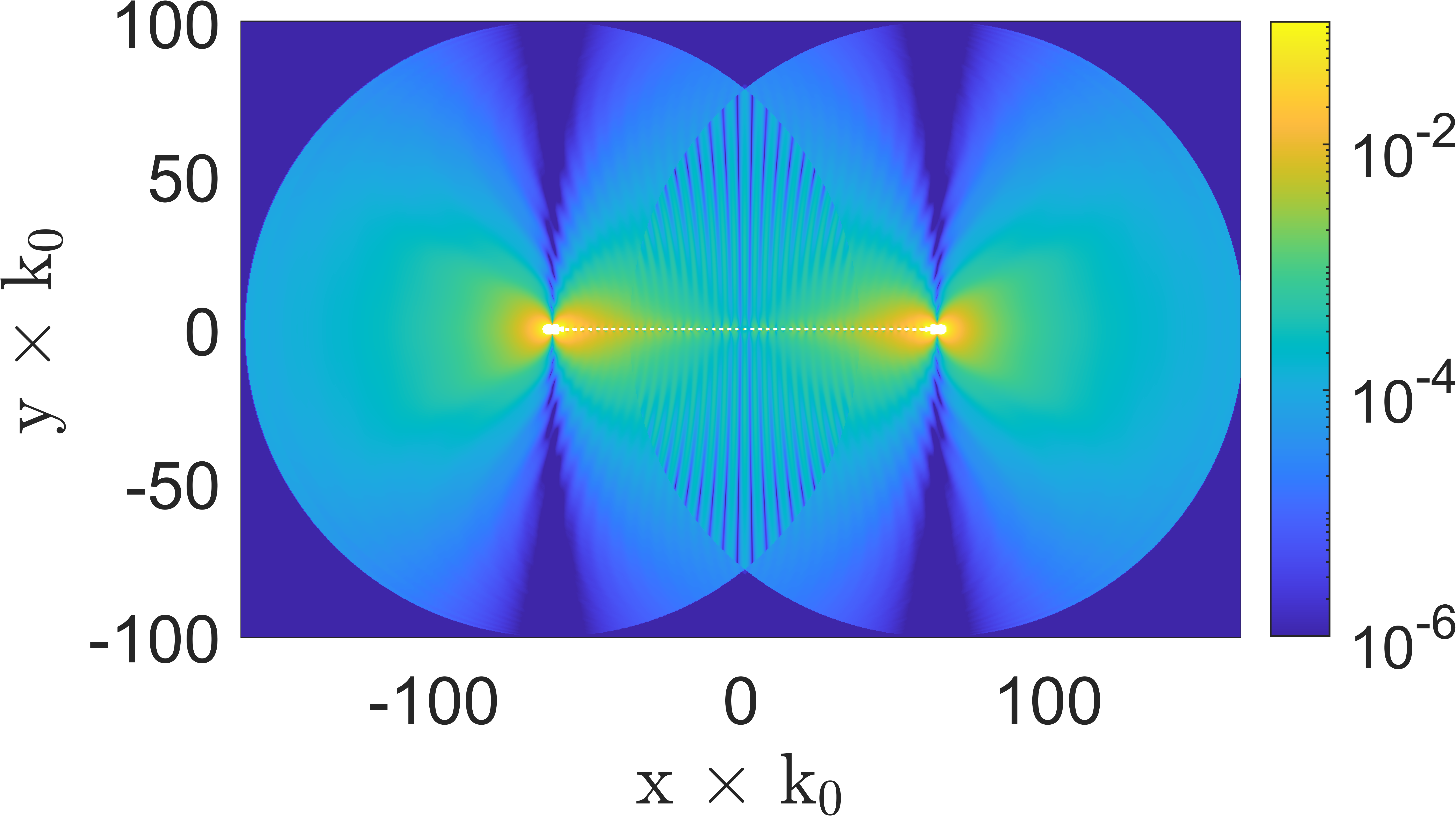}
            \put(0.000001,50){\small\bfseries f)}
        \end{overpic}
    \end{minipage}

    \caption{The near zone figures in the plane that atomic array lies; for $\theta=\pi/2$, $a=0.4 \lambda_0$, $u=0.8$, $t= 0.8 Na/c$ for the edge states and $t= 0.5 Na/c$ for the rest. The chain consists of $N=50$ unit cells. Panels (a--c) show emission from states with even unit-cell dipole-moment parity, while panels (d--f) show emission from states with odd parity. Panels (a,d) correspond to the two most superradiant states, panels (b,e) to the two most subradiant states, and panels (c,f) to the two edge states.Values above the upper color-scale threshold are masked to reveal the weaker near-zone spatial structure away from the chain. The threshold is chosen separately for each panel to improve visibility.}
    \label{fig:perpendicularnearfieldpatterns}

\end{figure*}

\begin{figure*}[!htbp]
    \centering

    \begin{minipage}[t]{0.325\textwidth}
        \begin{overpic}[width=\linewidth]{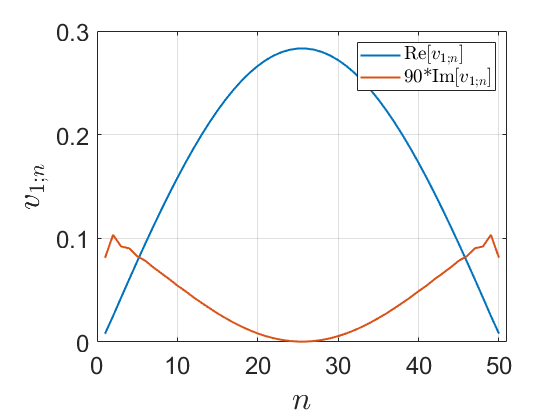}
            \put(0.0001,70){\small\bfseries a)}
        \end{overpic}
    \end{minipage}
    \hfill
    \begin{minipage}[t]{0.325\textwidth}
        \begin{overpic}[width=\linewidth]{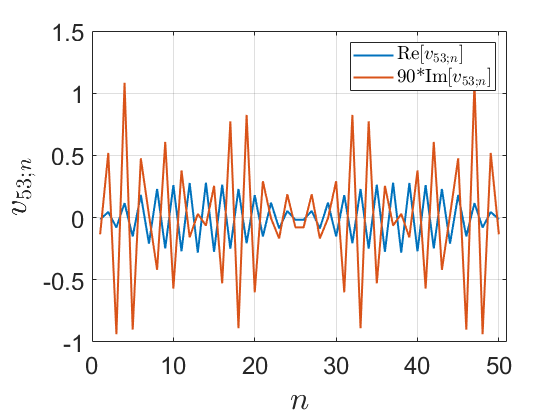}
            \put(0.00001,70){\small\bfseries b)}
        \end{overpic}
    \end{minipage}
    \hfill
    \begin{minipage}[t]{0.325\textwidth}
        \begin{overpic}[width=\linewidth]{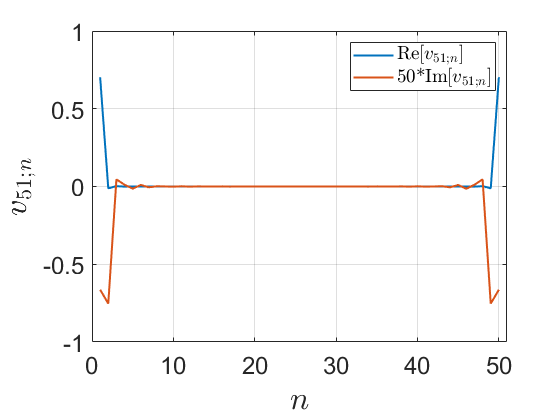}
            \put(0.0000001,70){\small\bfseries c)}
        \end{overpic}
    \end{minipage}


    \begin{minipage}[t]{0.325\textwidth}
        \begin{overpic}[width=\linewidth]{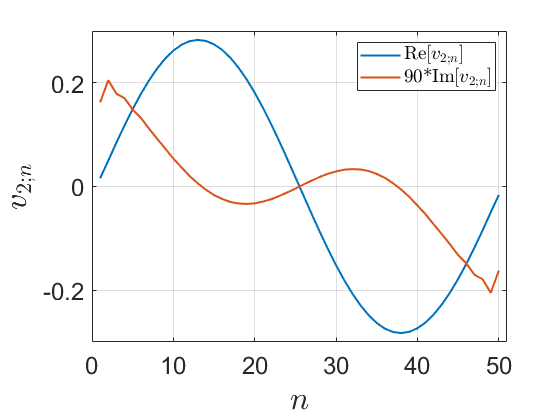}
            \put(0.0000001,70){\small\bfseries d)}
        \end{overpic}
    \end{minipage}
    \hfill
    \begin{minipage}[t]{0.325\textwidth}
        \begin{overpic}[width=\linewidth]{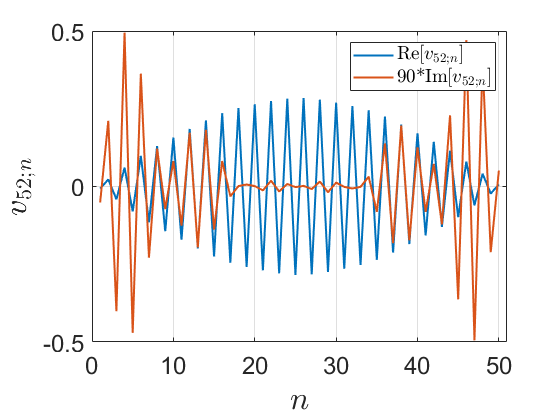}
            \put(0.0000001,70){\small\bfseries e)}
        \end{overpic}
    \end{minipage}
    \hfill
    \begin{minipage}[t]{0.325\textwidth}
        \begin{overpic}[width=\linewidth]{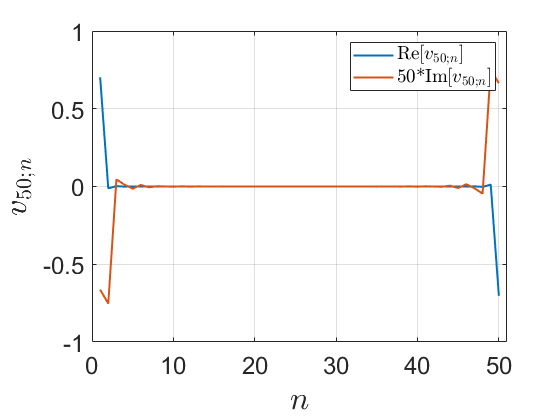}
            \put(0.000001,70){\small\bfseries f)}
        \end{overpic}
    \end{minipage}

    \caption{Unit cell amplitudes $v_{\xi; n}$ for different states; for $\theta=0$, $a=0.4 \lambda_0$, $u=0.8$. The chain consists of $N=50$ unit cells. The states correspond to the ones that are shown in radiation pattern figures. Panels (a--c) show states with even unit-cell dipole-moment parity, while panels (d--f) show states with odd parity. Panels (a,d) correspond to the two most superradiant states, panels (b,e) to the two most subradiant states, and panels (c,f) to the two edge states.}
    \label{fig:parallelparity}

\end{figure*}
\begin{figure*}[!htbp]
    \centering

    \begin{minipage}[t]{0.325\textwidth}
        \begin{overpic}[width=\linewidth]{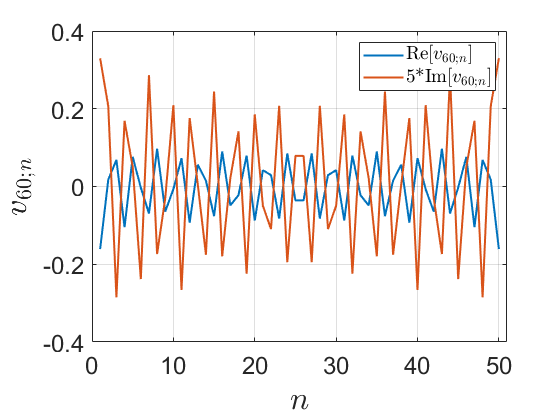}
            \put(0.0001,70){\small\bfseries a)}
        \end{overpic}
    \end{minipage}
    \hfill
    \begin{minipage}[t]{0.325\textwidth}
        \begin{overpic}[width=\linewidth]{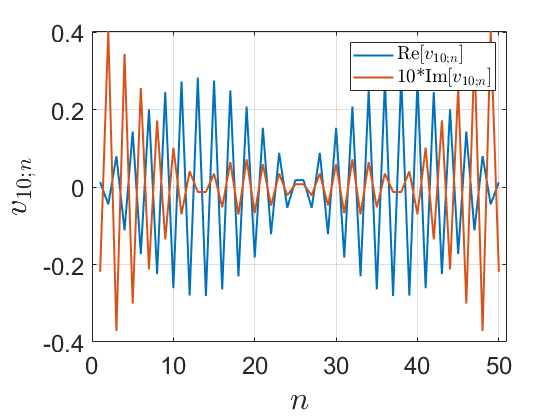}
            \put(0.00001,70){\small\bfseries b)}
        \end{overpic}
    \end{minipage}
    \hfill
    \begin{minipage}[t]{0.325\textwidth}
        \begin{overpic}[width=\linewidth]{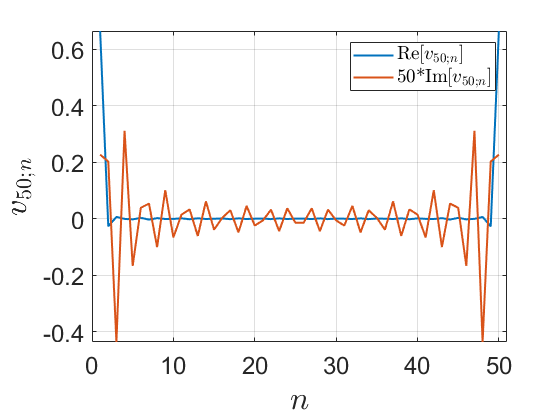}
            \put(0.0000001,70){\small\bfseries c)}
        \end{overpic}
    \end{minipage}


    \begin{minipage}[t]{0.325\textwidth}
        \begin{overpic}[width=\linewidth]{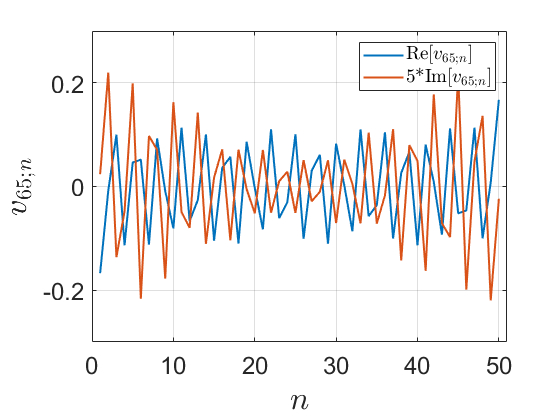}
            \put(0.0000001,70){\small\bfseries d)}
        \end{overpic}
    \end{minipage}
    \hfill
    \begin{minipage}[t]{0.325\textwidth}
        \begin{overpic}[width=\linewidth]{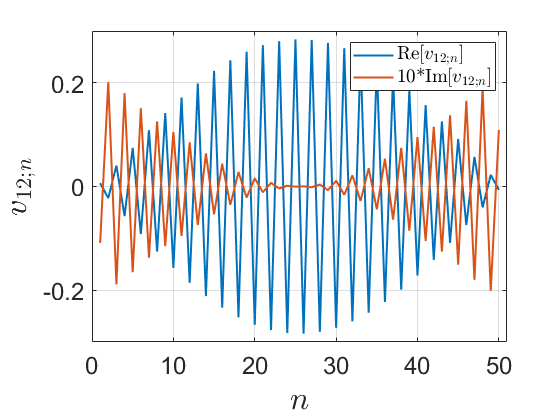}
            \put(0.0000001,70){\small\bfseries e)}
        \end{overpic}
    \end{minipage}
    \hfill
    \begin{minipage}[t]{0.325\textwidth}
        \begin{overpic}[width=\linewidth]{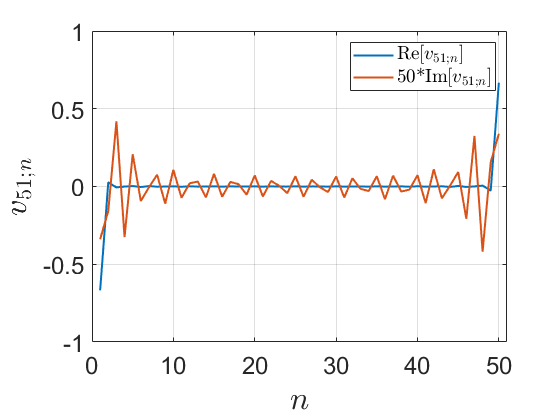}
            \put(0.000001,70){\small\bfseries f)}
        \end{overpic}
    \end{minipage}

    \caption{Unit cell amplitudes $v_{\xi; n}$ for different states; for $\theta=\pi/2$, $a=0.4 \lambda_0$, $u=0.8$. The chain consists of $N=50$ unit cells. The states correspond to the ones that are shown in radiation pattern figures. Panels (a--c) show states with even unit-cell dipole-moment parity, while panels (d--f) show states with odd parity. Panels (a,d) correspond to the two most superradiant states, panels (b,e) to the two most subradiant states, and panels (c,f) to the two edge states.}
    \label{fig:perpendicularparity}
\end{figure*}

The unit-cell amplitudes $v_{\xi,n} = \sum_\alpha  v_{\xi; n\alpha}$ are plotted in Fig.\ref{fig:parallelparity} and Fig.\ref{fig:perpendicularparity} for $\theta=0$ and $\pi/2$ respectively. The panels are in correspondence to the parameters and the states considered in Fig.\ref{fig:paraallelFarfieldpatterns}, Fig.\ref{fig:parallelnearfieldpatterns} and Fig.\ref{fig:Perpendicularfarfieldpatterns}, Fig.\ref{fig:perpendicularnearfieldpatterns}. In Fig.\ref{fig:parallelparity}(a--c) and Fig.\ref{fig:perpendicularparity}(a--c), the profiles have even parity with respect to the center of the chain, so the total unit-cell dipole moment is not symmetry-forbidden. By contrast, in Fig.\ref{fig:parallelparity}(d--f) and Fig.\ref{fig:perpendicularparity}(d--f) , the profiles have odd parity, leading to a vanishing total unit-cell dipole moment.  

The unit-cell dipole moment profiles provide a real-space indication of the radiative character of the collective eigenstates. The superradiant states in Fig.\ref{fig:parallelparity}(a,d) and Fig.\ref{fig:perpendicularparity}(a,d) display slowly varying amplitudes over the chain, corresponding to collective excitations with long characteristic wavelengths and small effective wave-vector components. Such modes are efficiently coupled to propagating radiation because the dipoles in different unit cells emit nearly in phase, resulting in constructive interference in the far field. By contrast, the subradiant states in Fig.\ref{fig:parallelparity}(b,e) and Fig.\ref{fig:perpendicularparity}(b,e) possess rapidly varying spatial profiles with alternating phases between nearby unit cells. This produces destructive interference of the emitted fields and strongly reduces the net radiative coupling. In reciprocal-space language, these states have dominant Fourier weight at larger wave vectors, which are poorly matched to the propagating modes of the electromagnetic field, especially when they lie outside or near the light line. Since the finite chain eigenstates are not exact Bloch waves, this wave-vector interpretation should be understood in terms of dominant Fourier components rather than sharply defined Bloch momenta. The edge states in Fig.\ref{fig:parallelparity}(c,f) and Fig.\ref{fig:perpendicularparity}(c,f) are instead characterized by their spatial localization at the boundaries rather than by an extended bulk wavelength.